\documentclass[12pt]{article}
\pdfoutput=1

\usepackage{adjustbox}
\usepackage{color}
\usepackage{tabularx,colortbl}
\usepackage{amssymb,amsmath,bm}
\usepackage{epsf}
\usepackage{epsfig}
\usepackage{afterpage}
\usepackage{longtable}
\usepackage[utf8]{inputenc}
\usepackage{cite}
\usepackage{latexsym,mathrsfs}
\usepackage{graphics}
\usepackage{parskip}
\usepackage{url}
\usepackage{paralist}
\usepackage{float}
\usepackage{booktabs}
\usepackage{multirow}
\usepackage[dvipsnames]{xcolor}
\usepackage[linktoc=page,bookmarks=false,colorlinks=false,linkbordercolor=RoyalBlue,citebordercolor=ForestGreen,urlbordercolor=CornflowerBlue]{hyperref}

\usepackage{caption}
\definecolor{green1}{rgb}{0.06,0.66,0.06}
\definecolor{orange1}{rgb}{0.98,0.60,0.07}
\definecolor{darkgreen}{rgb}{0.0,0.6,0.0}
\definecolor{darkblue}{RGB}{12,13,115}
\definecolor{darkred}{RGB}{204,6,0}

\newcommand{\gp}{$ {g_1}^2$}
\newcommand{\g}{$ g_2^2$}

\usepackage{titlesec}

\usepackage{fancyhdr}
\advance \headheight by 3.0truept       

\definecolor{red}{cmyk}{0,1,1,0.4}

\newcommand{\beq}{\begin{equation}}
\newcommand{\eeq}{\end{equation}}
\newcommand{\be}{\begin{equation}}
\newcommand{\ee}{\end{equation}}
\newcommand{\bi}{\begin{itemize}}
\newcommand{\ei}{\end{itemize}}
\newcommand{\ba}{\begin{array}}
\newcommand{\ea}{\end{array}}
\newcommand{\beqa}{\begin{eqnarray}}
\newcommand{\eeqa}{\end{eqnarray}}
\newcommand{\bea}{\begin{eqnarray}}
\newcommand{\eea}{\end{eqnarray}}
\newcommand{\beqn}{\begin{eqnarray}}
\newcommand{\eeqn}{\end{eqnarray}}

\newcommand{\nn}{\nonumber}

\newcounter{TODO}

\def \NP{{\rm NP}}

\def \refapp#1{Appendix~\ref{#1}}

\newcommand{\muEW}{{\mu_\text{ew}}}

\newcommand{\wc}[3][{}]{\big[{\cal C}_{#2}^{#1}\big]_{#3}}

\newcommand{\wclow}[3][{}]{\big[{ c}_{#2}^{#1}\big]_{#3}}

\newcommand{\op}[3][{}]{[{\cal O}_{#2}^{#1}]_{#3}}

\begin{document}


\bigskip
\begin{center}
{\LARGE\bf
\boldmath{Renormalization group improved implications of semileptonic operators in SMEFT}}
\\[0.8 cm]
{\bf Jacky Kumar
\\[0.5cm]}
{\small
TUM Institute for Advanced Study,
    Lichtenbergstr. 2a, D-85747 Garching, Germany}
\\[0.5 cm]
\footnotesize
E-Mail:
\texttt{jacky.kumar@tum.de}
\\[0.2 cm]

\end{center}
\vskip0.41cm
\abstract{%
\noindent 
We study implications of the four-fermion semileptonic 
operators at the low-energy and at electroweak (EW) scale in the framework of Standard Model Effective 
Field Theory (SMEFT). We show how the renormalization group (RG) running effects 
can play an important  role in probing the \emph{generic flavour structure} of such operators. 
It is shown that at the 1-loop level, through RG running, depending upon the flavour structure, 
these operators can give rise to sizable effects at low energy in the electroweak precision (EWP) observables, 
 the leptonic, quark, as well as the $Z$ boson flavour violating decays. To this end, we isolate the phenomenologically 
 relevant terms in the full anomalous dimension matrices (ADMs) and discuss the impact of  
the QED+QCD running in the Weak effective field theory (WET) and the SMEFT running due to gauge and 
Yukawa interactions on the dim-4 and dim-6 operators at the low energy.  Considering all the relevant processes,  
we derive lower bounds on new physics (NP) scale $\Lambda$ for each semileptonic operator, keeping a generic flavour 
structure. In addition, we also report the allowed ranges for the Wilson coefficients at a fixed value of $\Lambda=3$ TeV. 

}
\setcounter{page}{0}
\thispagestyle{empty}
\newpage
\setcounter{tocdepth}{4}
\tableofcontents
\newpage

\section{Introduction}
\label{sec:introduction}

In the absence of the discoveries of new particles at the Large Hadron Collider (LHC), 
the SMEFT provides an elegant framework to parameterize and quantify 
the effects of NP in terms of $SU(3)_c \times SU(2)_L \times U(1)_Y$ gauge invariant 
higher dimensional ($d\ge 5$) operators\cite{Grzadkowski:2010es, Buchmuller:1985jz}. 
In SMEFT, the operators are constructed using the field content of the Standard Model (SM). 
Excluding the flavour structures, there are in total 59 operators which conserve 
the baryon number\cite{Grzadkowski:2010es}.

An interesting aspect of the SMEFT is its built-in gauge symmetry. 
This feature often leads to an intriguing pattern of correlations among low energy observables due to 
 enforcement of the model independent relations between the Weak effective theory (WET) operators 
 at the EW scale, on matching with SMEFT \cite{Buras:2020xsm, Bhattacharya:2014wla, Capdevila:2017iqn, Buras:2014fpa}. 
 At the 1-loop level, through RG running, new effective operators can also be generated at the EW scale as 
a result of operator mixing. 
For instance, the four-fermion SMEFT operators can mix with the
 $\psi^2 \phi^2 D$ type operators which can contribute to the observables of different kinds. 
In this manner, the running from $\Lambda$ to the EW scale can induce additional correlations 
\cite{Feruglio:2017rjo, Cornella:2018tfd, Aebischer:2020mkv, Aebischer:2020lsx, Aebischer:2020dsw, Bruggisser:2021duo}.  
Therefore, in an SMEFT analysis, considering such effects is very important in order to correctly predict the 
low energy implications of new interactions introduced at the NP scale.

By now, the complete ADMs for SMEFT as well as WET are known at the 1-loop 
level\cite{Alonso:2013hga,Jenkins:2013wua,Jenkins:2013zja,Jenkins:2017dyc, Aebischer:2017gaw}.
The recent results for ADMs in the SMEFT extended with right-handed neutrino 
fields can be found in Refs.~\cite{Chala:2020pbn, Datta:2020ocb, Datta:2021akg}.
Based on these calculations, the tools such as {\tt wilson} \cite{Aebischer:2018bkb} and 
{\tt DsixTools} \cite{Celis:2017hod, Fuentes-Martin:2020zaz} have been developed. 
 (See also Ref.~\cite{Buras:2018gto} for a discussion on the analytic solutions to the SMEFT RGEs).    
Using these codes, it is possible to include the RG running effects in the theoretical predictions which can be obtained using
 {\tt flavio} \cite{Straub:2018kue}.
Several other packages \cite{Brivio:2017btx, Dedes:2019uzs, Brivio:2019irc} 
exist to facilitate the different kinds of tasks for studying the phenomenology in the SMEFT framework.  
As far as the present work is concerned, we have used {\tt flavio} for the theoretical predictions and 
the RG running effects have been taken care by using the {\tt wilson} package.

In this work, we focus on a subset of SMEFT operators which contain both lepton 
 and quark fields currents:
\begin{equation}
 (\overline L_i \gamma_\mu L_j) (\overline Q_k \gamma^\mu Q_l) \,, \notag
\end{equation}
here, \emph{i,j,k,l} denote the flavour indices. 
Such operators are known as the semileptonic operators. It is well-known that, at tree-level, these operators enter into 
various semileptonic decays of mesons. In this context, the operators which violate the 
quark flavour have been extensively studied in the literature
\cite{Cornella:2018tfd, Endo:2020kie, Aebischer:2020mkv, Alasfar:2020mne, Celis:2017doq, Kumar:2018kmr, Bobeth:2017ecx, Buttazzo:2017ixm, Calibbi:2015kma, Alonso:2015sja}.  
On the other hand, a generic flavour structure of these operators is not yet fully explored. 
In particular, the quark flavour conserving counterparts deserve more attention. 
A given NP model can generate both flavour violating as well as conserving operators, 
 therefore, it is essential to know what kind of constraints apply on the later ones.
One of the goals of the present work is to fill this gap by identifying all possible low energy 
and EW scale observables, which can be used to probe a generic flavour structure of the semileptonic operators. 
Concerning this matter, the Ref.~\cite{Dawson:2019clf} discusses the contribution 
of semileptonic operators to the EWP observables, assuming flavour universality.
Similarly, the Refs.~\cite{Feruglio:2016gvd, Feruglio:2017rjo}, pointed out the importance of EWP 
and lepton flavour violating (LFV)  constraints on the semileptonic operators needed to resolve the $B$-anomalies. 
For more recent studies on this topic, see also Refs.~\cite{Aebischer:2018iyb, Kumar:2018kmr, Alasfar:2020mne}.
Due to the reasons outlined above, we will restrict ourself to the flavour structures 
such that there is no quark-flavour violation, to begin with, at the scale $\Lambda$.
 In other words, the operators in which are interested are either  
flavour conserving in both currents or at most they violate the lepton flavour at the NP scale. 
We will emphasize the importance RG running from $\Lambda$ to the EW scale, through which these operators can contribute to a verity of observables at lower scales.
In this regard, we will first isolate the most important terms in the ADMs due to gauge couplings as well as Yukawas, 
\emph{i.e.}, the ones which are phenomenologically relevant, given the current precision of the measurements.

The remainder of the paper is structured as follows. In Sec.~\ref{sec:strategy}, we will discuss our strategy, 
 and in Sec.~\ref{sec:rgrunning}, we discuss the SMEFT RG running of semileptonic operators. In Sec.~\ref{sec:observables},  
  we will identify various observables which are relevant for the different flavour structures of the 
operators under consideration. 
In Sec.~\ref{sec:scales}, we will discuss the sensitivities to the NP scales $\Lambda$ for various operators. 
Finally, we move on to the conclusions in Sec.~\ref{sec:conclusion}. Additional material is collected in Appendices 
\ref{app:shiftsew}, \ref{app:comp-gauge-yuk}, and \ref{app:boundswcs}.

\section{General Strategy}
\label{sec:strategy}
In SMEFT, the SM is extended with a series of higher dimensional effective operators invariant 
under the full gauge symmetry of the SM. In general, the SMEFT Lagrangian can be written as
\begin{align}
\label{eq:1}
\mathcal{L}_{\rm SMEFT}^{d \ge 5} = \sum_{ \mathcal{ O}_a^\dagger = \mathcal{O}_a} \mathcal{C}_a \mathcal{O}_a 
+ \sum_{\mathcal{O}_a^\dagger \ne \mathcal{O}_a} \left ( \mathcal{C}_a \mathcal{O}_a + \mathcal{C}_a^* \mathcal{O}_a^\dagger  \right   )\,,
\end{align} 
here, $\mathcal{C}_a$ are known to be the Wilson coefficients. A complete list of SMEFT operators can be 
found in Refs.~\cite{Buchmuller:1985jz, Grzadkowski:2010es}. 
In this work, we will focus on a subset, the four-fermion semileptonic operators:
\begin{eqnarray} 
\label{eq:opsa}
\op[(1)]{\ell q}{ijkl} &=& (\bar \ell_i \gamma_\mu \ell_j)(\bar q_k \gamma^\mu q_l)\,,\\
\op[(3)]{\ell q}{ijkl} &=& (\bar \ell_i \gamma_\mu \tau^a \ell_j)(\bar q_k \gamma^\mu \tau^a q_l)\,,\\
\op[]{\ell u}{ijkl} &=& (\bar \ell_i \gamma_\mu \ell_j)(\bar u_k \gamma^\mu u_l)\,,\\
\op[]{\ell d}{ijkl} &=& (\bar \ell_i \gamma_\mu \ell_j)(\bar d_k \gamma^\mu d_l)\,,\\
\op[]{ed}{ijkl} &=& (\bar e_i \gamma_\mu e_j)(\bar d_k \gamma^\mu d_l)\,,\\
\op[]{eu}{ijkl} &=& (\bar e_i \gamma_\mu e_j)(\bar u_k \gamma^\mu u_l)\,,\\
\op[]{qe}{ijkl} &=& (\bar q_i \gamma_\mu q_j)(\bar e_k \gamma^\mu e_l)\,,
\label{eq:opsz}
\end{eqnarray}
here, the flavour indices $i,j,k,l$ can take values from 1 to 3 and $q, u, d, \ell, e$ represent the quark 
doublet, right-handed up-type quark, right-handed down-type quark, lepton doublet and right-handed 
lepton fields, respectively.  The down-type quarks are chosen to be in the mass basis at the scale $\Lambda$. 
This corresponds to the Warsaw-down basis of the SMEFT\cite{Aebischer:2017ugx}.
For convenience, we divide the operators into three classes based on their flavour structure at the scale $\Lambda$:
\begin{itemize}
\item $\Delta F= (0,0)$: the operators which do not violate quark\footnote{Since we work in the 
Warsaw-down basis, the Cabibbo–Kobayashi–Maskawa (CKM) rotations can give rise to quark flavour violating 
operators in the up-sector, even at the scale $\Lambda$. {Note that the CKM matrix itself can be 
affected by NP \cite{Descotes-Genon:2018foz, Aebischer:2018iyb}, however such effects are beyond the scope of 
present analyses. In our analyses for numerics the CKM is 
obtained with {\tt wilson} program using the input 
$V_{us}=0.2243, V_{ub}= 3.62 \times 10^{-3}, V_{cb}= 4.221 \times 10^{-2}, \delta=1.27$ at the $Z$-mass scale. }} and lepton flavours,
\item $\Delta F= (1,0)$: the operators which do violate lepton flavour by one unit but do not violate the quark flavour,
\item $\Delta F= (i,1)$: the operators which do violate quark flavour by one unit but may or may not violate the lepton flavour,
\emph{i.e}, $i= 0$ or 1.
\end{itemize}
Since, the $\Delta F= (i,1)$ type operators are already well studied in the literature, we will not consider them here.
On integrating out the new degrees of freedom at $\Lambda$, a unique tower of SMEFT operators is generated. 
However, a priori it is not obvious to which observables these operators can enter into at 
the lower energies. This is because of three reasons.
Firstly, the $SU(2)_L$ invariance can lead to correlations between different type of observables. 
Secondly, the pattern of mixing between different operators due to running from $\Lambda$ to the 
EW scale is very complex in nature. Often this leads to the appearance of new operators at the EW scale, which 
 can give rise to unpredictable correlations among low-energy observables.
Finally, the choice of the flavour basis at the NP scale is not invariant with respect to the RG evolution.
Therefore, it is extremely important to systematically analyze these effects for all flavour structures of the 
operators of our interest. This will allow us to identify all possible observables which are sensitive to these operators. 

With these motivations, first we will identify the most sensitive observables which can be used to probe the operators listed 
in \eqref{eq:opsa}-\eqref{eq:opsz} for the following flavour combinations:
\begin{eqnarray}
\label{eq:flavours1}
\Delta F=(0,0) &:&  1111, 1122, 1133, 2211, 2222, 2233, 3311, 3322, 3333\,,   \\
\Delta F= (1,0)&:&  1211, 1222, 1233, 1311, 1322, 1333, 2311, 2322, 2333.
\label{eq:flavours2}
\end{eqnarray}
Note that none of these flavour combinations are quark flavour violating. Next, we will proceed in 
three steps:
To begin with, we will study the operator mixing pattern due RG evolution for the two classes 
of the flavour structures as shown in Eqs.~\eqref{eq:flavours1}-\eqref{eq:flavours2}.  
Based on this, we will then identify all possible observables which can be used 
to constrain a given operator directly (at tree-level) or through the operators to which it mixes into,  
through the RG effects (at 1-loop level).
Using this information, we will finally derive the lower bounds on the scale of each 
semileptonic operator assuming the presence of a single operator at the scale $\Lambda$.

\section{Renormalization Group Running}
\label{sec:rgrunning}
In this section we discuss the RG running of the SMEFT semileptonic Wilson coefficients. 
In general, the running is governed by the coupled differential 
equations
\begin{equation}
\mathcal{\dot C}(\mu) \equiv 16\pi^2\frac{d\mathcal{C}(\mu)}{d \ln \mu} = \hat \gamma(\mu)
~\mathcal{C}(\mu)\,,
\end{equation}
with
\begin{equation}
\mathcal{C}(\mu)  = (\mathcal{C}_1(\mu), \mathcal{C}_2(\mu), ...)^T.
\end{equation}
Here, $\mu$ is the renormalization scale and $\hat \gamma$ is the anomalous dimension 
matrix which is function of the SM parameters such as gauge and Yukawa couplings. 
In the leading-log (LL) approximation, the solution to these equations for running from
scale $\Lambda$ and $\mu$ reads
\begin{equation}
\label{eq:llsolns}
\mathcal{C}(\mu) = \left [\hat 1 - \frac{\hat \gamma}{16 \pi^2} 
\ln \left ( \frac{\Lambda}{\mu}\right )\right ] \mathcal{C}(\Lambda).
\end{equation}

In the following we analyze the RG running and operator mixing of various quark-flavour 
conserving semileptonic operators (shown in Eqs.~\eqref{eq:opsa}-\eqref{eq:opsz})
of our interest to various other operators. These effects can potentially relate them to new observables 
generated at 1-loop level and hence allow us to put additional constraints.  
Using the ADMs calculated in 
Refs.~\cite{Alonso:2013hga,Jenkins:2013wua,Jenkins:2013zja,Jenkins:2017dyc} we find that, depending upon the 
flavour structures, the quark flavour conserving semileptonic operators can mix with two types of operators. 
The first category is the $\psi^2 \phi^2 D$ type operators:
\begin{eqnarray} 
\label{eq:rgeopsa}
\op[(1)]{\phi \ell}{ij} &=& (\phi^\dagger i \,{\overleftrightarrow {D_\mu}} \, \phi)(\bar{\ell}_i \, \gamma^\mu\, \ell_j) \,,\\
\op[(3)]{\phi \ell }{ij} &=&  (\phi^\dagger i\, {\overleftrightarrow {D_\mu^I}} \, \phi) (\bar \ell_i \tau^I \gamma^\mu \ell_j) \,,\\
\op[]{\phi e}{ij} &=& (\phi^\dagger i\, {\overleftrightarrow {D_\mu}} \, \phi) (\bar e_i  \gamma^\mu e_j) \,,\\
\op[(1)]{\phi q}{ij} &=&  (\phi^\dagger i\, {\overleftrightarrow {D_\mu}} \, \phi) (\bar q_i  \gamma^\mu q_j)   \,,\\
\op[(3)]{\phi q}{ij} &=& (\phi^\dagger i\, {\overleftrightarrow {D_\mu^I}} \, \phi) (\bar q_i \tau^I \gamma^\mu q_j)\,,\\
\op[]{\phi u}{ij} &=& (\phi^\dagger i\, {\overleftrightarrow {D_\mu}} \, \phi) (\bar u_i  \gamma^\mu u_j) \,, \\
\op[]{\phi d}{ij} &=& (\phi^\dagger i\, {\overleftrightarrow {D_\mu}} \, \phi) (\bar d_i \gamma^\mu d_j).
\label{eq:rgeopsz}
\end{eqnarray}
Here, $I$ is the $SU(2)$ index and $D_\mu$ stands for the covariant derivative.
Secondly, they also mix with the purely leptonic operators given by
\begin{eqnarray}
\label{eq:rgeopslla}
\op[]{\ell \ell}{ijkl} &=& (\bar \ell_i \gamma_\mu \ell_j)(\bar \ell_k \gamma^\mu \ell_l)\,, \\
\op[]{\ell e}{ijkl} &=& (\bar \ell_i \gamma_\mu \ell_j)(\bar e_k \gamma^\mu e_l)\,, \\
\op[]{ee}{ijkl} &=& (\bar e_i \gamma_\mu e_j)(\bar e_k \gamma^\mu e_l).
\label{eq:rgeopsllz}
\end{eqnarray}

{In addition, the $\Delta F=(0,0)$ type semileptonic operators can 
also mix with the four-quark operators, however we do not find any significant 
constraints due to such an operator mixing.}
In order to get the general picture, for our qualitative discussion, we will use the LL 
solutions to the corresponding RG equations {as given by Eq.~\eqref{eq:llsolns}}.
 However, for the numerics, we sum the logs using full numerical solutions which have been obtained using the 
{\tt wilson} program \cite{Aebischer:2018bkb}.

\bigskip

{\boldmath{
\subsection{Evolution of $\Delta F=(0,0)$ operators}}}
\noindent
The ADMs in the SMEFT depend on the gauge couplings as well the Yukawas. In this section, we identify 
the phenomenologically  important terms in the ADMs of $\Delta F=(0,0)$ semileptonic operators. 

\bigskip

\subsubsection{Operator mixing due to Gauge interactions}

First, we discuss the operator mixing of the $\Delta F=(0,0)$ semileptonic operators due the 
gauge couplings. The flavour combinations for these operators are 
specified in Eq.~\eqref{eq:flavours1}. Therefore, as an initial condition, 
at NP scale, only the Wilson coefficients of the operators given in 
\eqref{eq:opsa}-\eqref{eq:opsz} are assumed to be non-zero.
It turns out that, these operators can mix with the $\psi^2 \phi^2 D$-type  
operators, which are listed in Eqs.~\eqref{eq:rgeopsa}-\eqref{eq:rgeopsz}, through EW interactions. 
Now we use the LL approximation to relate the Wilson coefficients of the semileptonic operators at the scale $\Lambda$ 
to the Wilson coefficients of $\psi^2\phi^2 D$ operators which get generated at the EW scale due to the operator mixing. 
We find
\beq
                \begin{pmatrix}
                 \wc[(1)]{\phi \ell}{ii}\\
                 \wc[(3)]{\phi \ell}{ii}  \\
                \wc[]{\phi e}{ii} \\
                 \wc[(1)]{\phi q}{kk} \\
                \wc[(3)]{\phi q}{kk} \\
                 \wc[ ]{\phi d}{kk} \\
                 \wc[]{\phi u}{kk} 
                \end{pmatrix}  =
                L
                \begin{pmatrix}
                \frac{2}{3}$\gp$  & 0 & \frac{4}{3}$\gp$  & -\frac{2}{3}$\gp$ & 0   & 0 & 0\\
                0     & 2$\g$ &  0    &      & 0   & 0 & 0\\
                0     & 0 &  0    & 0    & -\frac{2}{3}$\gp$& \frac{4}{3}$\gp$ & \frac{2}{3}$\gp$ \\
                -\frac{2}{3}$\gp$    &  0 &  0    & 0    &  0  &  0  & -\frac{2}{3}$\gp$\\
                0    &  \frac{2}{3}$\g$ &  0    & 0    &  0  & 0 & 0\\
                0    &  0 & 0 & -\frac{2}{3}$\gp$  &  -\frac{2}{3}$\gp$  & 0 &0 \\
                0    &  0 &  -\frac{2}{3}$\gp$ & 0 &  0 &     -\frac{2}{3}$\gp$ & 0 \\
                \end{pmatrix}
                \begin{pmatrix}
                \wc[(1)]{\ell q}{iikk}\\
                \wc[(3)]{\ell q}{iikk}\\
                \wc[]{\ell u}{iikk}\\
                \wc[]{\ell d}{iikk}\\
                \wc[]{e d}{iikk}\\
                \wc[]{e u}{iikk}\\
                \wc[]{q e}{kkii}
        \end{pmatrix}.
\label{eq:rge-df00}
\eeq
Strictly speaking on the l.h.s. we should have written the difference between the Wilson coefficients 
 at the EW scale and their corresponding values at the scale $\Lambda$, \emph{i.e.,}
\begin{equation}
\delta \mathcal{C}({\muEW})= \mathcal{C}(\muEW)- \mathcal{C}(\Lambda).
\end{equation} 
However, since at the scale $\Lambda$ we assume only the semileptonic operators to be non-zero, we have 
$\mathcal{C}(\Lambda)=0$ and therefore $\delta \mathcal{C}({\muEW})= \mathcal{C}(\muEW)$.
Also, we have expressed the loop suppression factor and the log in terms of the quantity $L$, given by
\begin{equation} \label{eq:log}
L=\frac{1}{16\pi^2} \ln \left ( \frac{\muEW}{\Lambda} \right ).
\end{equation}
In Eq.~\eqref{eq:rge-df00}, the $g_1$ and $g_2$ are the EW gauge couplings at the scale  $\Lambda$. 
It is important to note that for the first three (last four) elements on the l.h.s., 
 only the repeated index $k$ ($i$) is summed over on the r.h.s. However, on the l.h.s. indices $k$ and $i$ are not 
summed over and can take values in the range 1-3. For simplicity, we do not show the self-mixing of the operators.
Interestingly, after the EW symmetry breaking, the $\psi^2 \phi^2 D$ operators are known to give 
corrections to the $W$ and $Z$ boson couplings with quarks and 
leptons \cite{Falkowski:2014tna, Ellis:2014jta, Efrati:2015eaa, Berthier:2015oma} 
(see also more recent  Refs. \cite{Aebischer:2018iyb,Ellis:2018gqa, Dawson:2019clf,  Breso-Pla:2021qoe, Falkowski:2019hvp, Brivio:2017vri, Dawson:2018pyl, Anisha:2020ggj}). Therefore, the quark-flavour 
conserving semileptonic operators can be indirectly probed through EWP data. 

In this context, one should also consider the operator $\wc[]{\ell \ell}{1221}$ which 
enters into the muon decay, \emph{i.e.}, $\mu \to e \nu \bar \nu$ and hence affects the extraction of Fermi constant 
$G_F$. At the EW scale the Wilson coefficient
 $\wc[]{\ell \ell}{1221}$ can be written as
\begin{equation}
\label{eq:ll1221}
\wc[]{\ell \ell}{1221}   =   4 g_2^2 L \wc[(3)]{\ell q}{iikk}\,, 
\end{equation}
here the indices on the r.h.s. are summed over $ii=11,22$ and $kk=11,22,33$.

In addition, we find the mixing between $\wc[(1)]{\ell q}{}$ and $\wc[(3)]{\ell q}{}$ to be phenomenologically relevant. 
At the 1-loop level, this goes through the EW corrections. Therefore, the strength of mixing is driven by the gauge 
coupling $g_2$. As before, solving the corresponding RGEs in LL approximation, one finds
\beq
                \begin{pmatrix}
                 \wc[(1)]{\ell q}{iikk}\\
                 \wc[(3)]{\ell q}{iikk}  
                \end{pmatrix}  =
                L
                \begin{pmatrix}
                 $--$     &  9$\g$  \\
                3$\g$     & $--$        \\
                \end{pmatrix}
                \begin{pmatrix}
                \wc[(1)]{\ell q}{iikk}\\
                \wc[(3)]{\ell q}{iikk}
        \end{pmatrix}\,,
\label{eq:rge-df00-2}
\eeq

here, the repeated indices $k$ and $i$ on the l.h.s. as well as on the r.h.s. are not summed over. 
Once again, we have suppressed the self-mixing (indicated by the entries with dashes) of these operators 
and log term $L$ is given by \eqref{eq:log}. The mixing of $\wc[(1)]{\ell q}{iikk}$ with $\wc[(3)]{\ell q}{iikk}$ 
can induce new charged current transitions after 
EW symmetry breaking. 

\subsubsection{Operator mixing due to top-Yukawa interactions}

In this subsection, we isolate phenomenologically important terms in the ADMs which depend on the 
 Yukawa interactions. In this regard, we will keep only the largest terms involving the top-Yukawa 
coupling. Solving the RGEs in the LL approximation, we find 
\beq
                \begin{pmatrix}
                 \wc[(1)]{\phi  \ell}{ij}\\
                 \wc[(3)]{\phi \ell}{ij} \\
                 \wc[ ]{\phi  e}{ij}
                \end{pmatrix}  \simeq
                6 y_t^2L
                \begin{pmatrix}
                 1 ~   &   0~   &  -1~ & 0~ & 0~ \\
                 0 ~   &   -1~  &   0~ & 0~ & 0~ \\
                 0 ~   &    0~  &   0~ & 1~ & -1 ~ \\
                \end{pmatrix}
                \begin{pmatrix}
                \wc[(1)]{\ell q}{ij33}\\
                \wc[(3)]{\ell q}{ij33}\\
                \wc[   ]{\ell u}{ij33}  \\
                \wc[   ]{ qe }{33ij}  \\
                \wc[   ]{ e u}{ij33}  
        \end{pmatrix}\,,
\label{eq:rge-df00-df10-yuk}
\eeq
here, $y_t$ is the Yukawa coupling for the top-quark.
In Eq.~\eqref{eq:rge-df00-df10-yuk}, for $\Delta F=(0,0)$ operators the flavour indices 
$i$ and $j$ can be set equal. It is worth mentioning that for simplicity, we have not included the non-standard effects 
due to running of the Yukawa couplings themselves, which can lead to additional contributions to the quark-flavour violating semileptonic 
operators at the EW scale. We will return to this point in Sec.~\ref{sec:bdecays}.

\bigskip

{\boldmath{
\subsection{Evolution of $\Delta F=(1,0)$ operators}}}
Now we consider the $\Delta F=(1,0)$ type semileptonic operators which violate the leptonic flavour by one unit 
but conserve the quark flavour (see Eq.~\eqref{eq:flavours2}). 
Apart from the self-mixing, these operators mix with the lepton flavour violating leptonic and $\psi^2\phi^2 D$ type operators,  
 $\wc[(1)]{\phi \ell}{ij}, \wc[(1)]{\phi \ell}{ij}$ and $\wc[]{\phi e}{ij}$ with $i \ne j$,  
 through the EW interactions. Once again, solving the RG equations in the LL approximation, we find,
\beq
                \begin{pmatrix}
                \wc[]{\ell \ell}{ijll}\\
                 \wc[]{\ell e}{ijll}  \\
                \wc[]{\ell e}{llij}  \\
                \wc[]{e e}{ijll} \\
                \wc[(1)]{\phi \ell}{ij} \\
                 \wc[(3)]{\phi \ell}{ij} \\
                 \wc[ ]{\phi e}{ij} \\
                \end{pmatrix} =
                L
                \begin{pmatrix}
                -\frac{1}{3}$\gp$  & $\g$ & -\frac{2}{3}$\gp$  & \frac{1}{3}$\gp$ & 0   & 0 & 0\\
                -\frac{4}{3}$\gp$  & 0 &  -\frac{8}{3}$\gp$    & \frac{4}{3}$\gp$     & 0   & 0 & 0\\
                 0     & 0 &  0    & 0    & \frac{2}{3}$\gp$& -\frac{4}{3}$\gp$ & -\frac{2}{3}$\gp$ \\
                0    &  0 &  0    & 0    &  \frac{2}{3}$\gp$  & -\frac{2}{3}$\gp$ & -\frac{1}{3}$\gp$\\
                \frac{2}{3}$\gp$  & 0 & \frac{4}{3}$\gp$  & -\frac{2}{3}$\gp$ & 0   & 0 & 0\\ 
                0     & 2$\g$ &  0    &      & 0   & 0 & 0\\ 
                0     & 0 &  0    & 0    & -\frac{2}{3}$\gp$& \frac{4}{3}$\gp$ & \frac{2}{3}$\gp$ \\ 
                \end{pmatrix}
                \begin{pmatrix}
                \wc[(1)]{\ell q}{ijkk}\\
                \wc[(3)]{\ell q}{ijkk}\\
                \wc[]{\ell u}{ijkk}\\
                \wc[]{\ell d}{ijkk}\\
                \wc[]{e d}{ijkk}\\
                \wc[]{e u}{ijkk}\\
                \wc[]{q e}{kkij}\\
        \end{pmatrix}\,,
\label{eq:rge-df10}
\eeq
here the repeated indices $k$ on the r.h.s. is summed over the values $1-3$, and
$i \ne j$ holds on the both sides. Also, the repeated index $l$ on the l.h.s. can take values in the range $1-3$ but 
it is not summed over. 
The logarithmic piece $L$ is defined in equation \eqref{eq:log}. 
The operators on the l.h.s. contribute to the LFV $W$, $Z$ boson couplings and leptonic decays.  
We will come to this point later with more details.
Note that, below the EW scale, the semileptonic operators can mix into purely leptonic operators 
 also through QED interaction \cite{Crivellin:2017rmk, Cirigliano:2021img}.

\bigskip

\section{Observables induced at 1-loop level}
\label{sec:observables}

Based on our discussions in the previous section, we are now in position to identify 
 the 1-loop induced observables to which the semileptonic operators can contribute. 
 We will see how the SMEFT RG running can play an important role in this respect. 
As discussed before, because of the complicated operator mixing pattern in SMEFT, 
the semileptonic operators can contribute to observables of very different nature  
depending upon the operators with which they mix, which can be read out from the l.h.s. of 
Eqs.~\eqref{eq:rge-df00}, \eqref{eq:ll1221}, \eqref{eq:rge-df00-2}, \eqref{eq:rge-df00-df10-yuk} 
and \eqref{eq:rge-df10}. In order to get a general picture, we will 
 present the expressions for the relevant low energy dim-4 and dim-6 couplings in terms of semileptonic 
 Wilson coefficients at the high  scale by employing the LL approximation.
 Eventually, we will sum the logs with the help of numerical solutions.

{\boldmath{
\subsection{Observables for $\Delta F=(0,0)$ operators}}}
\label{sec:obs-df00}
We have identified three categories of the observables which are relevant for $\Delta F=(0,0)$ type operators. 
Those are the EWP observables, flavour violating $B$-decays and charged current decays. 
In the following, we discuss how various semileptonic operators can contribute to them through operator mixing.

\subsubsection{Electroweak Precision Observables}
\label{sec:ewpobs}

As indicated by Eq.~\eqref{eq:rge-df00}, the $\Delta F=(0,0)$ semileptonic 
operators can mix with the $\psi^2 \phi^2 D$ type operators due to electroweak interactions.
In addition, this mixing also depends on the top-Yukawa interactions, 
see e.g. Eq.~\eqref{eq:rge-df00-df10-yuk}.  
Interestingly, after EW symmetry breaking the latter operators give corrections to the $Z$ and $W$ 
boson couplings with the fermions. Using the LL solutions to the RGEs, as presented 
before, we can express the NP contributions to these couplings directly in terms of the Wilson coefficients 
of semileptonic operators at the high scale $\Lambda$. In general, NP shifts in the neutral $Z$ boson 
couplings with the fermions can be parameterized as
\begin{equation}
\label{eq:def-delta}
\mathcal{L}_{Z } \supset g_Z \sum_{\psi =u, d, e, \nu_L} \bar \psi_i \gamma^\mu 
	\left [ \delta (g_L^\psi )_{ij} P_L 
	+ \delta (g_R^\psi )_{ij}  P_R  \right ] \psi_j Z_\mu\,,
\end{equation}
here $g_Z = - g_2 / \cos {\theta_W} $ and $\theta_W$ represents the weak-mixing angle. 
  NP can enter into $\delta g_X^\psi$ with $X=L,R$, through three difference sources. This can be understood from 
the equation\cite{Brivio:2017vri}
\begin{equation}
\label{eq:zsmnp}
\delta (g_X^{\psi})_{ij} =   \delta g_Z ~(g_X^{\psi, {\rm SM}})_{ij} - Q_\psi ~ \delta \sin^2 \theta_W~ \delta_{ij}  
+ \delta (g_X^{\psi})_{ij}^{\rm dir}\,,
\end{equation}
here, the contributions in the first two and the last term can be thought of as indirect and direct 
shifts to the $Z$ boson couplings, respectively.
In SMEFT, the tree-level expressions for 
 the quantities $\delta g_Z$, $\delta \sin^2 \theta_W$ and $\delta (g_X^\psi)_{ij}$ can be found in 
Appendix \ref{app:shiftsew}. Also, $Q_\psi$ represents the electric charge and 
 $g_L^{\psi, {\rm SM}} = T_3- Q_\psi \sin^2 \theta_W$, $g_R^{\psi, {\rm SM}}= -Q_\psi \sin^2 \theta_W$.  
Below we present the 1-loop contributions to all these quantities due to semileptonic operators in the LL approximation. 
Using the Eqs.~\eqref{eq:rge-df00} and \eqref{eq:zcouplingsa}-\eqref{eq:zcouplingsz} 
one can obtain the NP shifts in the $Z$ couplings with quarks:
\begin{eqnarray}
\label{eq:zcoup1loopa}
	\delta (g_L^u)_{kk}^{\rm dir} &=& -\frac{g_1^2}{3} v^2 L 
V_{km}  \left ( -\wc[(1)]{\ell q}{iimn} - \frac{g_2^2}{g_1^2} \wc[(3)]{\ell q}{iimn} -  \wc[]{qe}{mnii} \right ) V_{nk}^\dagger \,,\\
	\delta (g_R^u)_{kk}^{\rm dir} &=& -\frac{g_1^2}{3} v^2 L 
\left ( -\wc[(1)]{\ell u}{iikk} - \wc[(3)]{e u}{iikk} \right )\,,\\
	\delta (g_L^d)_{kk}^{\rm dir} &=& -\frac{g_1^2}{3} v^2 L 
\left ( -\wc[(1)]{\ell q}{iikk} + \frac{g_2^2}{g_1^2} \wc[(3)]{\ell q}{iikk} -   \wc[]{qe}{kkii} \right ) \,,\\
      \delta (g_R^d)_{kk}^{\rm dir} &=& -\frac{g_1^2}{3} v^2 L 
\left ( -\wc[]{\ell d}{iikk} - \wc[]{ed}{iikk}  \right ).
\end{eqnarray}
Similarly, the shifts in the $Z$ boson couplings with leptons are found to be
\begin{eqnarray}
	\delta (g_L^\nu)_{ii}^{\rm dir} &=& -\frac{g_1^2}{3} v^2 L 
\left ( \wc[(1)]{\ell q}{iikk} - \frac{3g_2^2}{g_1^2} \wc[(3)]{\ell q}{iikk} + 2\wc[]{\ell u}{iikk} - \wc[]{\ell d}{iikk}  \right ) \notag \\
&-&  6 v^2 y_t^2 L \left (\wc[(1)]{\ell q}{ii33} +\wc[(3)]{\ell q}{ii33} - \wc[]{\ell u}{ii33} \right ) \,, \\
	\delta (g_L^e)_{ii}^{\rm dir} &=& -\frac{g_1^2}{3} v^2 L 
\left ( \wc[(1)]{\ell q}{iikk} + \frac{3g_2^2}{g_1^2} \wc[(3)]{\ell q}{iikk} + 2\wc[]{\ell u}{iikk} - \wc[]{\ell d}{iikk}  \right ) \notag \\
&-&  6 v^2 y_t^2 L \left (\wc[(1)]{\ell q}{ii33} -\wc[(3)]{\ell q}{ii33}  - \wc[]{\ell u}{ii33} \right ) \,, \\
	\delta (g_R^e)_{ii}^{\rm dir} &=& -\frac{g_1^2}{3} v^2 L 
\left ( -\wc[]{e d}{iikk} + 2 \wc[ ]{eu}{iikk} + \wc[]{qe}{kkii} \right ) \notag \\
&-&  6 v^2 y_t^2 L \left (\wc[]{qe}{33ii} - \wc[]{ eu}{ii33}  \right ). 
\label{eq:zcoup1loopz}
\end{eqnarray}
Here, $v=246$ GeV is the vacuum expectation value and $L$ is given by \eqref{eq:log}.
Similarly, for $\delta g_Z$ and $\delta \sin^2 \theta_W$ are found to be
\begin{eqnarray}
\label{eq:gz-1loop}
\delta g_Z &= &    3 v^2 y_t^2 L \left (\wc[(3)]{\ell q}{1133} +\wc[(3)]{\ell q}{2233} \right )     \,, \\ 
\delta \sin^2 \theta_W &= &    - \frac{\sin^2 2\theta_W}{ 4 \cos 2 \theta_W}   
6 v^2 y_t^2 L  \left (\wc[(3)]{\ell q}{1133} +\wc[(3)]{\ell q}{2233} \right ).
\label{eq:s2w-1loop}
\end{eqnarray}
On the basis of the above discussions, now we point out a few important observations:

\begin{itemize}
\item First, note that the RG induced EW corrections to $\delta g_Z$ and $\delta \sin^2 \theta_W$
 due to $\wc[(3)]{\ell q}{iikk}$ for $ii=11,22$ and $kk=11,22,33$, get canceled among 
$\wc[(3)]{\phi \ell}{11}, \wc[(3)]{\phi \ell }{22}$ 
and $\wc[]{\ell \ell}{1221}$. This can be seen by inserting the LL contributions 
of $\wc[(3)]{\ell q}{iikk}$  from Eqs.~\eqref{eq:ll1221} and \eqref{eq:rge-df00} 
to the latter Wilson coefficients in Eqs.~\eqref{eq:gztree}-\eqref{eq:s2wtree}.
However, this is no longer true once the contribution due to the Yukawa couplings, 
as shown in Eq.~\eqref{eq:rge-df00-df10-yuk}, is included.
\item  Next, in general, the top-Yukawa dependent contributions, given in 
Eqs.\eqref{eq:gz-1loop}-\eqref{eq:s2w-1loop}, are 
 larger in size as compared to the direct EW corrections to the $Z$-couplings 
 shown in Eqs.~\eqref{eq:zcoup1loopa}-\eqref{eq:zcoup1loopz}. So, for the cases in which both effects 
exist simultaneously, the former has a greater impact.
\item Furthermore, it is evident from Eq.~\eqref{eq:zsmnp} that the impact of the 
 $\wc[(3)]{\ell q}{ii33}$ for $ii=11$ or $22$ on $\delta (g_X^Y)_{ii}$ 
through $\delta g_Z$ and $\delta \sin^2 \theta_W$ 
is universal for all three families of leptons and quarks. 
On the other hand the shifts $\delta (g_{L}^e)_{ii}^{\rm dir}$, $\delta (g_{R}^e)_{ii}^{\rm dir}$  
and $\delta (g_L^\nu)_{ii}^{\rm dir}$ also experience effects due to 
the top-Yukawa interactions, which however are 
lepton flavour dependent. 
\item Finally, the contributions of the semileptonic operators due to top-Yukawa effects do not affect the 
$Z$ boson couplings to quarks directly. This is however still possible through $\delta g_Z$ and $\delta \sin^2 \theta_W$, but only for the case of $\wc[(3)]{\ell q}{ii33}$ with $ii=11,22$.
\end{itemize}
Next, we look at the impact of semileptonic operators on the $W$-boson couplings which can be parameterized as 
\begin{equation}
\label{eq:wcoup}
\mathcal{L}_{W} \supset \frac{-e}{\sqrt{2} \sin \theta_W}  
\left ( \delta (\varepsilon_L^\ell)_{ij}^{\rm dir}~ \bar \nu_i \gamma_\mu P_L e_j   +
\delta (\varepsilon_L^q)_{ij}^{\rm dir}~ \bar u_i \gamma_\mu P_L d_j  \right ) W^\mu_+ +h.c. ~\,,
\end{equation}
here the NP shifts are given as\cite{Brivio:2017vri}
\begin{eqnarray}
\delta (\varepsilon_L^\psi){ii} &=& \delta (\varepsilon_L^\psi)_{ii}^{\rm dir}  
-\frac{1}{2} \frac{\delta \sin^2 \theta_W}{ \sin^2\theta_W}. 
\end{eqnarray}
Using the LL solutions, we find the shifts 
$\delta (\varepsilon_L^\psi)_{ii}^{\rm dir}$ in terms of the semileptonic 
operators to be 
\begin{eqnarray}
\delta (\varepsilon_L^\ell)_{ii}^{\rm dir} &=&  2 g_2^2 L \wc[(3)]{\ell q}{iikk}  - 6 v^2 y_t^2 L \wc[(3)]{\ell q}{ii33}\,,\\ 
\delta (\varepsilon_L^q)_{kk}^{\rm dir} &=&  \frac{2}{3} g_2^2 L V_{km}  \wc[(3)]{\ell q}{iimk}.
\end{eqnarray}
We make the following observations for the RG induced shifts in the $W$ couplings:
\begin{itemize}
\item The $W$ couplings to both quarks and leptons are universally affected for all three families 
by $\wc[(3)]{\ell q}{ii33}$ 
for $ii=11,22$ through $\delta \sin^2 \theta_W$.
\item The top-Yukawa effects do not give direct contributions to the $W$ couplings. 
\item The leptonic couplings can be directly affected by  $\wc[(3)]{\ell q}{ii33}$ for $ii=11,22,33$ through top-Yukawa interactions. 
This effect is however flavour dependent.
\end{itemize}
In order to quantify these effects, in Appendix \ref{app:comp-gauge-yuk} we report 
tables for the numerical values of the RG induced shifts in the $Z$ and $W$ couplings at the EW scale due to  
semileptonic operators present at the NP scale $\Lambda$. To understand the relative importance  
of the top-Yukawa and gauge interactions, we present two sets of numbers with and without Yukawa RG running effects. 
It is evident that, whenever present, the top-Yukawa effects always dominate.

Now, given that the $Z$ and $W$ boson couplings are strongly 
constrained by the EWP observables, this implies that the flavour conserving semileptonic $\Delta F=(0,0)$ type 
operators can be indirectly constrained by the EWP measurements. The following list of operators can be 
constrained through this mechanism:

\begin{equation}
\label{eq:df00ewp}
\wc[(1)]{\ell q}{ijkl}\,, \wc[(3)]{\ell q}{ijkl}\,,  \wc[]{ ed }{ijkl}\,, \wc[]{ eu}{ijkl}\,, 
\wc[]{\ell u}{ijkl}\,, \wc[]{\ell d}{ijkl} \,, \wc[]{qe}{ijkl}\,, 
\end{equation}
with

\begin{equation}
ijkl \equiv 1111,1122,1133,2211,2222,2233,3311,3322,3333.
\end{equation}
In the next section, we will use the EWP measurements as constraints to derive the lower bounds on the NP scales of 
the flavour conserving semileptonic operators in given Eq.~\eqref{eq:df00ewp}. To this end, the list of EWP observables 
\cite{D0:1999bqi, LHCb:2016zpq, ALEPH:2013dgf, CDF:2013dpa, ATLAS:2017rzl, SLD:2000jop, ALEPH:2005ab} 
used in our analysis are taken from Tab.~13 of Ref.~\cite{Aebischer:2018iyb}. In addition, we have also 
included a recent measurement of the ratio 
$R_{\tau \mu} (W\to \ell \bar \nu)= \mathcal{B}(W\to \tau \bar \nu)/\mathcal{B}(W\to \mu \bar \nu) 
$\cite{ATLAS:2020csn}.

{
\boldmath{
\subsubsection{$B$ Meson Decays}
}}
\label{sec:bdecays}
\noindent
Throughout the analyses, we have used Warsaw-down basis at the high scale. In this basis, to begin with the down-type 
quark and lepton mass matrices are diagonal, whereas, the up-type quark matrix takes the form 
$V^\dagger {\rm diag}(y_u,y_c,y_t)$. Here, $V$ represents the CKM matrix.  
However, due to the RG running of the Yukawa matrices, this choice of basis is not preserved with respect to 
running from $\Lambda$ to the EW scale. 
As a result, one has to perform back-rotation to the original (Warsaw-down) basis at the EW scale. 
This process can generate flavour violating semileptonic operators from their flavour conserving counterparts at the scale $\Lambda$ 
\cite{Aebischer:2020lsx, Coy:2019rfr, Aebischer:2020mkv, Aebischer:2018bkb}.
The part of the WET Lagrangian that describes the $b \to s \ell^+ \ell^-$ processes can be written as
\bea
\mathcal{L}_{\rm eff} &=&  \frac{\alpha G_F}{\sqrt{2} \pi} V_{tb} V_{ts}^*
      \sum_{a = 9,10} ( C_a O_a + C'_a O'_a ) ~, \nn\\
O_{9(10)}^{ij} &=& [ {\bar s} \gamma_\mu P_L b ] [ {\bar e_i} \gamma^\mu (\gamma_5) e_j ]\,, \,\
O_{9(10)}^{\prime ij} = [ {\bar s} \gamma_\mu P_R b ] [ {\bar e_i} \gamma^\mu (\gamma_5) e_j ].
\label{Heff}
\eea
We can match them to the corresponding SMEFT Wilson coefficients as\cite{Aebischer:2015fzz} :
\bea
C_{9,\NP}^{ij} &=& \frac{\pi}{\alpha} \frac{v^2}{V_{tb} V_{ts}^*}
\left ( \wc[(1)]{\ell q}{ij23} + \wc[(3)]{\ell q}{ij23} + \wc[]{qe}{23ij} \right )\,, \nn \\
C_{10,\NP}^{ij} &=& \frac{\pi}{\alpha} \frac{v^2}{V_{tb} V_{ts}^*}
\left( \wc[]{qe}{23ij} - \wc[(1)]{\ell q}{ij23} - \wc[(3)]{\ell q}{ij23} \right)\,, \nn \\
C_{9,\NP}^{\prime ij} &=& \frac{\pi}{\alpha} \frac{v^2}{V_{tb} V_{ts}^*}
\left( \wc[]{\ell d}{ij23} + \wc[]{ed}{ij23} \right)\,, \nn\\
C_{10,\NP}^{\prime ij} &=& \frac{\pi}{\alpha} \frac{v^2}{V_{tb} V_{ts}^*}
\left( \wc[]{ed}{ij23} - \wc[]{\ell d}{ij23} \right ) ~.
\eea
Here, $ij = \mu\mu, ee$. We find that, through back-rotation,  the following semileptonic 
operators can contribute to $b\to s \ell^+ \ell^-$ observables:
\begin{equation}
\label{eq:opsBanomalies}
\wc[(1)]{\ell q}{ijkl}\,, \wc[]{ed}{ijkl}\,, \wc[]{\ell d}{ijkl}\,, \wc[]{qe}{klij}\,,
\end{equation}
with
\begin{equation}
i,j,k,l \equiv 1122, 1133, 2222, 2233.
\end{equation}
To constrain these operators we have used the measurements for LFUV observable such as 
$R_{K^{(*)}}$ \cite{Hiller:2003js, LHCb:2014vgu,LHCb:2019hip ,BELLE:2019xld, LHCb:2021trn, Aaij:2017vbb, Belle:2019oag} and 
$Q_{4,5}=P_{4,5}^{\mu \prime}- P_{4,5}^{e \prime}$ \cite{Belle:2016fev, Capdevila:2016ivx}. 
However, the full $b\to s \ell^+ \ell^-$ data set can give rise to stronger 
constraints\cite{Datta:2019zca, Altmannshofer:2021qrr}.

\bigskip

\subsubsection{Charged Current Decays}
\noindent
The $\rm (V-A) \times (V-A)$ type charged-current (c.c) interactions in low energy WET are 
given by effective Lagrangian
\begin{equation}
\mathcal{L}_{\rm eff}^{\rm c.c} = [C_{\nu edu}^{V,LL}]_{ijkl} ~ (\bar \nu_i \gamma_\mu P_L  e_j)(\bar d_k \gamma_\mu P_L u_l)  + h.c.~ .
\end{equation}
After the EW symmetry breaking, $[C_{\nu edu}^{V,LL}]_{ijkl}$ can be matched with the SMEFT operator $\wc[(3)]{\ell q}{ijkl}$. 
Now at 1-loop level, using \eqref{eq:rge-df00-2}, one can express $[C_{\nu edu}^{V,LL}]_{ijkl}$ in terms of 
 high scale Wilson coefficient $\wc[(1)]{\ell q}{ijkl}$ due to its mixing with $ \wc[(3)]{\ell q}{ijkl}$. 
We find
\begin{equation}
\label{eq:ccdecays}
[C_{\nu edu}^{V,LL}]_{ijkl} = {6 g_2^2} L   \wc[(1)]{\ell q}{ijkm} (V^\dagger)_{ml} ~.
\end{equation}
Here, the CKM elements $(V^\dagger)_{ml}$ are needed to rotate the up-quark states to the mass basis.
The above equation indicates that through EW corrections the singlet operator $\wc[(1)]{\ell q}{ijkl}$ can lead to 
 charged current transitions at the low energy. In the following we will look at several charged current processes which can be 
affected through this mechanism.

\bigskip

\paragraph{\boldmath $K$ Meson Decays} \mbox{}\\
\smallskip

\noindent
At the quark level, the charged-current $K$ decays involve the transition $s \to u e \bar \nu$. Therefore, these
 decays are driven by the WET operator $(\bar \nu_1 \gamma_\mu P_L e_1) (\bar d_2 \gamma^\mu P_L u_1)$. 
Now, using Eq.~\eqref{eq:ccdecays}, we note that this operator can be generated at the 1-loop level from the 
high scale SMEFT Wilson coefficient $\wc[(1)]{\ell q}{1122}$, \emph{i.e.},
\begin{equation}
[C_{\nu edu}^{V,LL}]_{1121} \propto \left ( V_{ud}^* \wc[(1)]{\ell q}{1121} + V_{us}^* \wc[(1)]{\ell q}{1122} + 
V_{ub}^* \wc[(1)]{\ell q}{1123}   \right ).
\end{equation}
Therefore, at the low-energy, $\wc[(1)]{\ell q}{1122}$ can lead to effects in the 
$K_L \to \pi e\bar \nu$, $K_S\to \pi e\bar \nu$, $K^+\to \pi e \bar \nu$, 
 and $K^+ \to \ell \bar \nu$ decays.

\paragraph{\boldmath $\tau \to K  \nu$ and $\tau \to \pi \nu$ \bf Decays} \mbox{}\\
\smallskip

\noindent
The $\tau \to K  \nu$ and  $\tau \to \pi \nu$ decays involve
$\tau  \to \bar u s \nu$ and $\tau  \to \bar  u d \nu$ transitions, respectively. 
At the low energy these are governed by the WET operators  
 $(\bar \nu_3 \gamma_\mu P_L e_3) (\bar d_2 \gamma^\mu P_L u_1)$ and  
$(\bar \nu_3 \gamma_\mu P_L e_3) (\bar d_1 \gamma^\mu P_L u_1)$. However, at the EW scale these operators can be 
generated from high scale Wilson coefficients $\wc[(1)]{\ell q}{3322}$ and $\wc[(1)]{\ell q}{3311}$ 
respectively (see \eqref{eq:ccdecays}), because
\begin{eqnarray}
{[C_{\nu edu}^{V,LL}]}_{3321} &\propto& \left ( V_{ud}^* \wc[(1)]{\ell q}{3321} + V_{us}^* \wc[(1)]{\ell q}{3322} +  V_{ub}^* \wc[(1)]{\ell q}{3323}   \right )\,, \\
{[C_{\nu edu}^{V,LL}]}_{3311} &\propto& \left ( V_{ud}^* \wc[(1)]{\ell q}{3311} + V_{us}^* \wc[(1)]{\ell q}{3312} + V_{ub}^* \wc[(1)]{\ell q}{3313}   \right ).
\end{eqnarray}
Therefore, one can use hadronic $\tau$ decays to constrain the SMEFT operators on the r.h.s.

\paragraph{\boldmath \bf $\pi \to e \nu$ and Nuclear $\beta$ Decays} \mbox{}\\
\label{sec:betadecay}
\smallskip

\noindent
The nuclear $\beta$-decay and $\pi \to e \nu$ are controlled by the WET operator
 $(\bar \nu_1 \gamma_\mu P_L e_1) (\bar d_1 \gamma^\mu P_L u_1)$. From \eqref{eq:ccdecays}, clearly at 
the EW scale this operator can be generated from $\wc[(1)]{\ell q}{1111}$ because
\begin{equation}
[C_{\nu edu}^{V,LL}]_{1111} \propto \left ( V_{ud}^* \wc[(1)]{\ell q}{1111} + V_{us}^* \wc[(1)]{\ell q}{1112} + 
V_{ub}^* \wc[(1)]{\ell q}{1113}   \right ).
\end{equation}
To summarize, we have found that the following list of high scale coefficients can contribute to various charged current 
decays at low energy:
\begin{equation}
\wc[(1)]{\ell q}{1122}\,, \wc[(1)]{\ell q}{3322}\,, \wc[(1)]{\ell q}{3311}\,, \wc[(1)]{\ell q}{1111}.
\end{equation}
Note that the Wilson coefficient $\wc[(3)]{\ell q}{ijkl}$  also give rise to tree-level contributions to the charged 
current decays. 
The experimental measurements for various charged current decays are shown in 
Tab.~\ref{tab:ccexp}. In addition, the measurements for the $\beta$-decays are taken 
from the Ref.~\cite{Hardy:2014qxa}.
\begin{table}[!htb]
  \begin{center}
\begin{adjustbox}{width=0.85\textwidth}
\begin{tabular}{|lr|lr|}
\toprule
Observable &  Experimental value & Observable & Experimental value \\ \hline 
$\mathcal{B}(K_L \to \pi e \bar \nu)$ & $ (40.55 \pm 0.11) \times 10^{-2} $ & $\mathcal{B}(K_S\to \pi e \bar \nu)$ & $(7.04 \pm 0.08) \times 10^{-4}$  \\
$\mathcal{B}(K^+ \to \pi e \bar \nu)$ & $ (5.07 \pm 0.04) \times 10^{-2} $  & $R_{e/\mu}({K^+ \to \ell \bar \nu})$ & $ (2.488 \pm 0.009) \times 10^{-5}$ \\ \hline
$\mathcal{B}(\tau \to K \nu)$ & $(6.96\pm 0.10)\times 10^{-3}$ & $\mathcal{B}(\tau \to \pi  \nu)$ & $(10.82\pm 0.05)\times 10^{-2}$ \\ \hline
$\mathcal{B}(\pi \to e \nu)$ & $(1.2344 \pm 0.0023 \pm 0.0019)\times 10^{-4} $  & & \\
\hline
\end{tabular}
\end{adjustbox}
\caption{\small The experimental measured values for the charged current  $K$, $\pi$ and $\tau$ 
decays \cite{ParticleDataGroup:2014cgo, PiENu:2015seu}.}
\label{tab:ccexp}
\end{center}
\end{table}

{Note that for the charged current operators, 
in principle the three-body $\tau$ decays such as
 the Belle spectrum for the process $\tau^- \to K_S \pi^- \nu_\tau$ also apply \cite{Belle:2007goc, Pich:2013lsa}. 
But as shown in Ref. \cite{Gonzalez-Solis:2020jlh} these decays 
give rise to similar constraints as $\tau^-\to K  \nu$ decay.}
\subsubsection{Correlations}
\begin{figure}[tb]
        \begin{center}
        \includegraphics[width=0.45\textwidth]{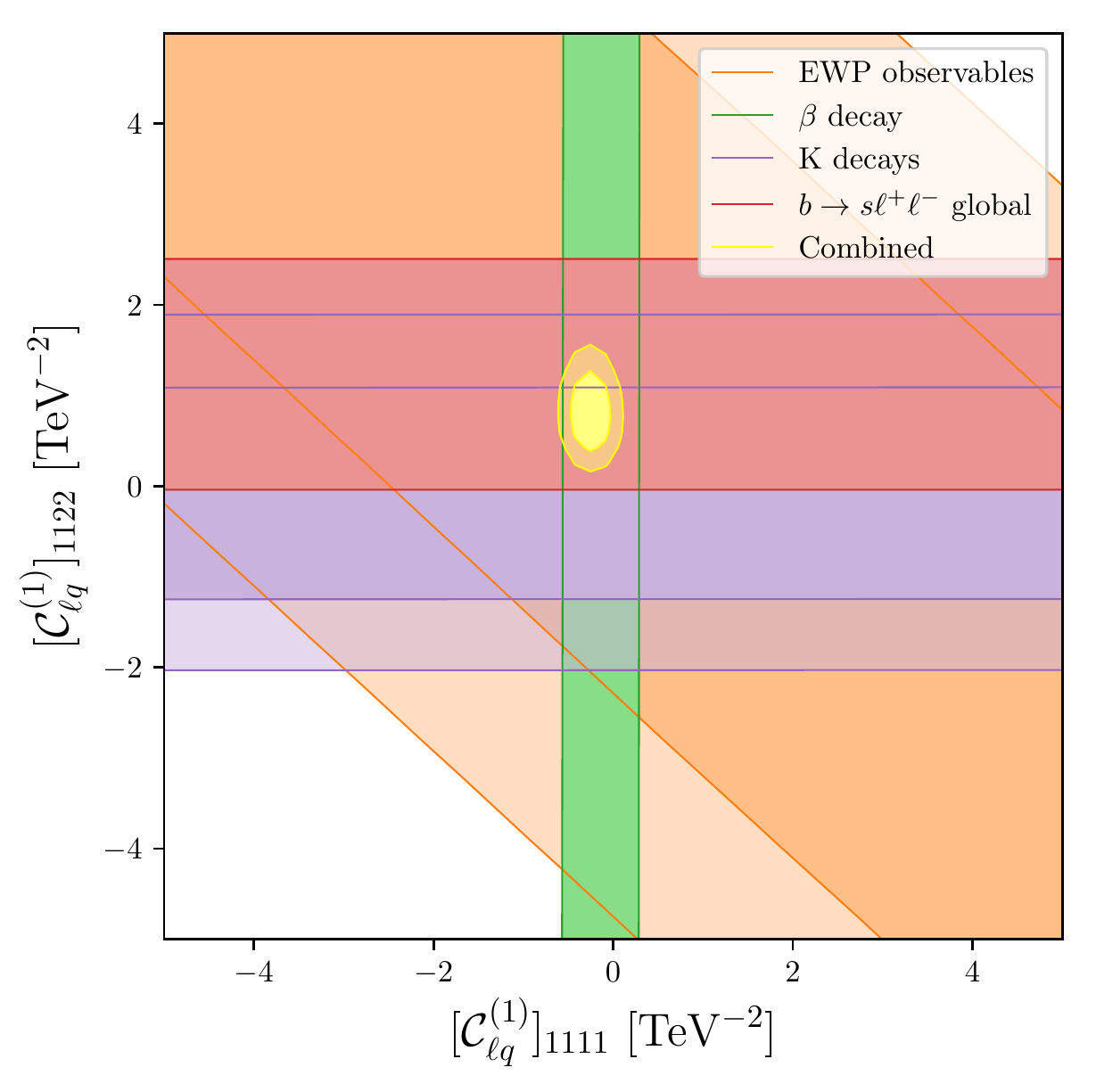}
        \includegraphics[width=0.45\textwidth]{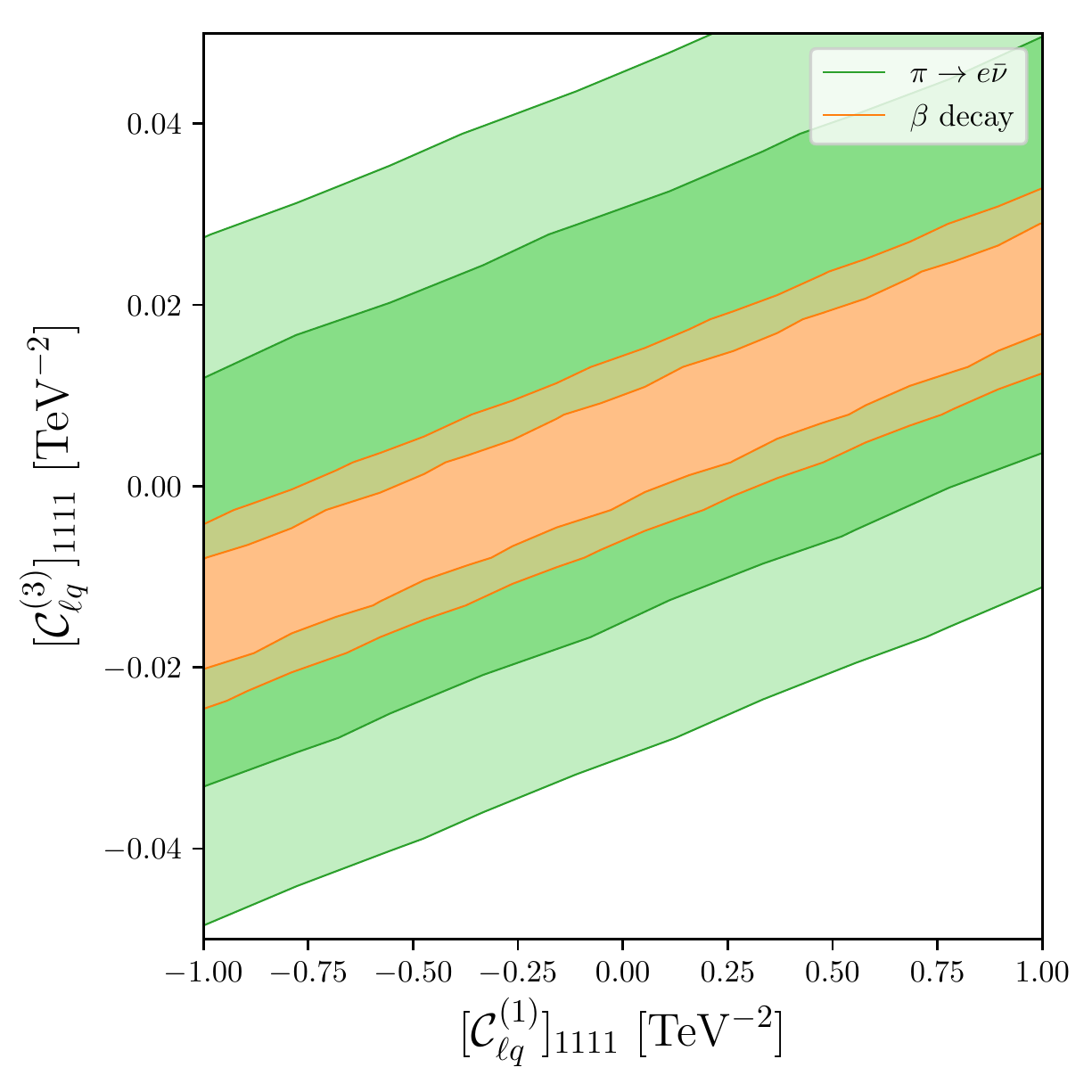}
        \end{center}
        \caption{\small The correlations between various observables in the 
        $\wc[(1)]{\ell q}{1111}$ - $\wc[(1)]{\ell q}{1122}$~ plane (left) and  
        $\wc[(1)]{\ell q}{1111}$ - $\wc[(3)]{\ell q}{1111}$~ plane (right) .}
        \label{fig:corr1}
\end{figure}

Since a single operator can contribute to several different kinds of observables through 
RG running, it would be interesting to see how these are correlated to each 
other. For instance, the operator $\wc[(1)]{\ell q}{1111}$ can contribute 
to $\beta$ decay as well as EWP observables. Similarly, $\wc[(1)]{\ell q}{1122}$ 
contributes to EWP observables, $b\to s \ell^+ \ell^-$ , as well as charged 
current $K$ processes. Take another example of the operator $\wc[(3)]{\ell q}{1111}$, which 
in addition to $\beta$ decay also contributes to $\pi \to e \bar \nu$ process at tree level. 
In Fig.~\ref{fig:corr1}, we show correlations between constraints due to 
various observables in the $\wc[(1)]{\ell q}{1111}$ - $\wc[(1)]{\ell q}{1122}$ and 
$\wc[(1)]{\ell q}{1111}$ - $\wc[(3)]{\ell q}{1111}$ planes in left and right panels respectively.  
Clearly, in order to get a complete picture about the constrains on a given operator, 
it is very important to take into account all RG induced observables.  

{\boldmath{
\subsection{Observables for $\Delta F=(1,0)$ operators}}}
\bigskip
Next, we move on to the semileptonic operators which violate the lepton flavour by one unit and conserve 
the quark flavour at the scale $\Lambda$. In this case, depending upon the Dirac and flavour structures, both tree-level as 
well as the 1-loop generated LFV observables are found to be important. 
The 1-loop generated processes are $\ell_i \to \ell_j \bar \ell \ell$ and $Z\to \bar \ell_i \ell_j$ 
 LFV decays. These processes are found to be relevant for all $\Delta F=(1,0)$ operators under consideration.
In addition, the $\Delta F=(1,0)$ semileptonic operators can also contribute to
 $\tau \to P \ell$ for $P=\pi, \phi$ and $\tau \to \rho \ell $ processes at the tree-level.

\bigskip

{\boldmath
\subsubsection{$Z\to \bar \ell_i \ell_j$ Decays}
}

\bigskip

The tree-level SMEFT contributions to the LFV $Z$ boson couplings  due to $\psi^2\phi^2 D$ operators 
  are given in Eqs.~\eqref{eq:zcouplingsa}-\eqref{eq:zcouplingsz}.
 At the 1-loop level, the semileptonic $\Delta F=(1,0)$ operators can mix with these $\psi^2 \phi^2 D $-type
operators, as shown in the Eqs.~\eqref{eq:rge-df10} and \eqref{eq:rge-df00-df10-yuk}. 
As a result the $\Delta F=(1,0)$ semileptonic operators can be constrained by the $Z\to \bar \ell_i \ell_j$ LFV decays.
The current experimental limits on these decays are given in Tab.~\ref{tab:lfvexp}.

\bigskip

{\boldmath
\subsubsection{$\tau \to 3 \ell$ and $\mu \to 3 e$ Decays}
}

In WET, the LFV processes such as $\tau \to 3 \ell$ and $\mu \to 3 e$ are governed by purely leptonic 
operators, as given by the effective Lagrangian:
\begin{eqnarray}
        \mathcal{L}_{\rm eff}^{\rm 1loop} &=& 
        [C_{ee}^{V,LL}]_{ijll}  ~(\bar e_{i} \gamma^\mu P_L e_{j}) (\bar e_{l} P_L \gamma_\mu e_{l}) \notag \\
        &+&    [C_{ee}^{V,LR}]_{ijll} ~(\bar e_{i} \gamma^\mu P_L e_{j}) (\bar e_{l} \gamma_\mu P_R e_{l}) \notag \\
        &+&    [C_{ee}^{V,LR}]_{llij} ~(\bar e_{l} \gamma^\mu P_L e_{l}) (\bar e_{i} \gamma_\mu P_R e_{j}) \notag \\
        &+&    [C_{ee}^{V,RR}]_{ijll} ~(\bar e_{i} \gamma^\mu P_R e_{j}) (\bar e_{l} \gamma_\mu P_R e_{l}) + h.c. ~.
\end{eqnarray}

Here, the superscript indicates that such operators are generated only at the 1-loop 
level and are set to zero at the NP scale.
In SMEFT, at tree-level the corresponding four fermion leptonic operators 
are $\wc[]{\ell \ell}{ijkl}$, $\wc[]{\ell e}{ijkl}$, and $\wc[]{ee}{ijkl}$ 
 which are defined in Eqs.~\eqref{eq:rgeopslla}-\eqref{eq:rgeopsllz}. In addition to this, the SMEFT can also 
contribute through $\psi^2\phi^2 D$ type operators after integrating out the $Z$ boson.  
Such contributions arise by combining the LFV effective $Z$ boson vertices due to SMEFT with the flavour conserving 
$Z$ boson interactions in the SM\cite{Jenkins:2017jig}. The $\Delta F=(1,0)$ semileptonic operators can give rise to 
both effects through operator mixing.  From Eq.~\eqref{eq:rge-df10}, 
one can find that, at the 1-loop level through the EW interactions, 
 the contributing four-fermion SMEFT operators can be directly generated from the semileptonic operators. 
In addition, the $\psi^2 \phi^2 D$ operators get contributions through both the 
gauge \eqref{eq:rge-df10} as well as the top-Yukawa 
interactions \eqref{eq:rge-df00-df10-yuk}.
In the LL approximation, combining the four fermion and $\psi^2 \phi^2 D$ type contributions, 
we can express the contributing WET Wilson coefficients at the EW scale directly in terms of SMEFT 
Wilson coefficients of the semileptonic operators at $\Lambda$:

\begin{eqnarray}
{[C_{ee}^{V,LL}]}_{ijll}  &=&  
\frac{g_1^2}{3} L  \left ( -  \wc[(1)]{\ell q}{ijkk} 
+  \frac{3 g_2^2}{g_1^2} \wc[(3)]{\ell q}{ijkk} - 2  \wc[]{\ell u}{ijkk} +   \wc[]{\ell d}{ijkk}  \right )\notag     \\   
&+& \frac{\mathcal{Z}}{4}  \left ( \wc[(1)]{\ell q}{ij33} -  \wc[(3)]{\ell q}{ij33} 
- \wc[ ]{\ell u}{ij33}   \right )  g_L^{e, {\rm SM}}    \,, \\
{[C_{ee}^{V,LR}]}_{ijll}  &=&  \frac{4 g_1^2}{3} L 
\left (   - \wc[(1)]{\ell q}{ijkk}  - 2 \wc[]{\ell u}{ijkk} +   \wc[]{\ell d}{ijkk}  \right ) \notag \\
&+&  \mathcal{Z} \left ( \wc[(1)]{\ell q}{ij33} -  \wc[(3)]{\ell q}{ij33} 
- \wc[ ]{\ell u}{ij33}   \right )  g_R^{e, {\rm SM}}    \,, \\
{[C_{ee}^{V,LR}]}_{llij}  &=&  \frac{2g_1^2}{3} L 
\left (    \wc[(1)]{ed}{ijkk}  - 2  \wc[]{e u}{ijkk} -   \wc[]{qe}{kkij}  \right ) \notag    \\     
&+& \mathcal{Z}  \left ( \wc[(1)]{qe}{33ij} -  \wc[]{eu}{ij33}  \right )  g_L^{e, {\rm SM}}    \,, \\
{[C_{ee}^{V,RR}]}_{ijll}  &=&  \frac{2 g_1^2}{3} L
\left (    \wc[(1)]{ed}{ijkk} -  \wc[]{e u}{ijkk} -   \wc[]{qe}{kkij}  \right ) \notag \\
&+& \frac{\mathcal{Z}}{4}  \left ( \wc[(1)]{qe}{33ij} -  \wc[]{eu}{ij33}  \right )  g_R^{e, {\rm SM}}    \,, 
\end{eqnarray}
with the factor $\mathcal{Z}  = \left [  3 v^2 y_t^2 L \right ]  \left [\frac{g_Z^2}{M_Z^2} \right ]$.  
Here $L$ is the log term defined in \eqref{eq:log}.
Numerically, the comparison of the influence of the gauge and Yukawa interactions on the WET Wilson coefficients on the l.h.s. 
is presented in Tab.~\ref{tab:wetleptonic} in Appendix \ref{app:rgesemilep} .
We find that the following set of Wilson coefficients can be constrained through this mechanism:
\begin{equation}
\wc[(1)]{\ell q}{ijkl}\,, \wc[(3)]{\ell q}{ijkl}\,,  \wc[]{ ed }{ijkl}\,, \wc[]{ eu}{ijkl}\,, 
\wc[]{\ell u}{ijkl}\,, \wc[]{\ell d}{ijkl} \,, \wc[]{qe}{klij}\,, 
\end{equation}
with
\begin{equation}
ijkl \equiv 1211, 1222, 1233, 1311, 1322, 1333, 2311, 2322, 2333.
\end{equation}
The experimental limits used for the LFV decays of $\tau$  and $\mu$ 
leptons are collected in Tab.~\ref{tab:lfvexp}.

\bigskip

{\boldmath
\subsubsection{$\tau \to \rho \ell$,  $\tau \to P(\phi, \pi) \ell$ Decays}
}
\noindent
In the WET,  the hadronic decays $\tau \to P \ell$ for $P=\pi,\phi$ and $\tau \to \rho \ell$ can be described 
 by the effective Lagrangian

\begin{eqnarray}
	\mathcal{L}_{\rm eff}^{\rm tree+1loop} &=& 
	\wc[V,LL]{ed}{ijkk} (\bar e_{i} \gamma^\mu P_L e_{j}) (\bar d_{k} \gamma_\mu P_L d_{k}) +
	\wc[V,LL]{eu}{ij11} (\bar e_{i} \gamma^\mu P_L e_{j}) (\bar u_{1}  \gamma_\mu P_L u_{1}) \notag  \\
	&+&\wc[V,LR]{ed}{ijkk} (\bar e_{i} \gamma^\mu P_L e_{j}) (\bar d_{k} \gamma_\mu P_R d_{k}) +
	\wc[V,LR]{eu}{ij11} (\bar e_{i} \gamma^\mu P_L e_{j}) (\bar u_{1} \gamma_\mu P_R u_{1}) \notag \\
	&+&\wc[V,LR]{de}{kkij} (\bar d_{k} \gamma^\mu P_L d_{k}) (\bar e_{i} \gamma_\mu P_R e_{j}) +
	\wc[V,LR]{ue}{11ij} (\bar u_{1} \gamma^\mu P_L u_{1}) (\bar e_{i} \gamma_\mu P_R e_{j}) \notag \\ 
	&+&\wc[V,RR]{ed}{ijkk} (\bar e_{i} \gamma^\mu P_R e_{j}) (\bar d_{k} \gamma_\mu P_R d_{k}) +
	\wc[V,RR]{eu}{ij11} (\bar e_{i} \gamma^\mu P_R e_{j}) (\bar u_{1} \gamma_\mu P_R u_{1}). \notag \\
& & 
\end{eqnarray}
Here, the indices $kk=11$ for $\tau \to \pi \ell$, $\tau \to \rho \ell$ and $22$ 
for $\tau \to \phi \ell$. In addition to the tree-level contributions, the semileptonic operators in SMEFT 
can also contribute to these operators through top-Yukawa loops. Adding the both contributions, in the 
LL approximation, at the EW scale we obtain:

\begin{eqnarray}
\label{eq:VLLedijkk}
	\wc[V,LL]{ed}{ijkk} &=& \wc[(1)]{\ell q}{ijkk} + \wc[(3)]{\ell q}{ijkk}  \notag\\
&+& \mathcal{Z}  \left [ \wc[(1)]{\ell q}{ij33} -  \wc[(3)]{\ell q}{ij33} 
- \wc[ ]{\ell u}{ij33}   \right ]  g_L^{d, {\rm SM}}    \,, \\
\label{eq:VLLeuijkk}
	\wc[V,LL]{eu}{ij11} &=& V_{1m} \left [\wc[(1)]{\ell q}{ijmn} - \wc[(3)]{\ell q}{ijmn} \right ] V^\dagger_{n1}  \notag \\
&+&  \mathcal{Z} \left [ \wc[(1)]{\ell q}{ij33} -  \wc[(3)]{\ell q}{ij33} 
- \wc[ ]{\ell u}{ij33}   \right ]  g_L^{u, {\rm SM}}     \,,  \\
	\wc[V,LR]{ed}{ijkk} &=& \wc[]{\ell d}{ijkk} 
~+~ \mathcal{Z}  \left [ \wc[(1)]{\ell q}{ij33} -  \wc[(3)]{\ell q}{ij33} 
- \wc[ ]{\ell u}{ij33}   \right ] g_R^{d, {\rm SM}}    \,, \\  
	\wc[V,LR]{eu}{ij11} &=&  \wc[]{\ell u}{ij11}
~+~ \mathcal{Z} \left [ \wc[(1)]{\ell q}{ij33} -  \wc[(3)]{\ell q}{ij33} 
- \wc[ ]{\ell u}{ij33}   \right ]  g_R^{u, {\rm SM}}    \,, \\  
	\wc[V,LR]{de}{kkij} &=& \wc[]{qe}{kkij}
~+~ \mathcal{Z} \left [ \wc[]{qe}{33ij} -  \wc[]{eu}{ij33}   \right ]  g_L^{d, {\rm SM}}     \,, \\  
	\wc[V,LR]{ue}{11ij} &=& V_{1m}  \wc[]{qe}{mnij} V^\dagger_{n1}  
~+~ \mathcal{Z} \left [ \wc[]{qe}{33ij} -  \wc[]{eu}{ij33}    \right ]  g_L^{u, {\rm SM}}     \,, \\  
	\wc[V,RR]{ed}{ijkk} &=& \wc[]{ed}{ijkk} 
~+~ \mathcal{Z} \left [ \wc[]{qe}{33ij} -  \wc[]{eu}{ij33}   \right ]  g_R^{d, {\rm SM}}     \,, \\  
	\wc[V,RR]{eu}{ij11} &=& \wc[]{eu}{ij11} 
~+~ \mathcal{Z} \left [ \wc[]{qe}{33ij} -  \wc[]{eu}{ij33}   \right ]  g_R^{u, {\rm SM}} .
\end{eqnarray}
Here, the CKM elements $V_{ij}$ on the r.h.s. are needed to rotate the up-quarks from the Warsaw-down to 
 the mass basis for matching them with the WET Wilson coefficients on the l.h.s. 
The quantity $\mathcal{Z} = \left [  3 v^2 y_t^2 L \right ]  \left [\frac{g_Z^2}{M_Z^2} \right ]$. 
Numerically, the relative influence of the WET and SMEFT running can be found in 
Tab.~\ref{tab:wethadronic}. The following set of Wilson coefficients can be 
constrained through this mechanism:
\begin{equation}
\wc[(1)]{\ell q}{ijkl}\,, \wc[(3)]{\ell q}{ijkl}\,,  \wc[]{ ed }{ijkl}\,, \wc[]{ eu}{ijkl}\,,
\wc[]{\ell u}{ijkl}\,, \wc[]{\ell d}{ijkl} \,, \wc[]{qe}{klij}\,,
\end{equation}
with
\begin{equation}
ijkl \equiv 1211, 1222, 1233, 1311, 1322, 1333, 2311, 2322, 2333.
\end{equation}
Note that the operators involving $u\bar u$, $d\bar d$ and $s \bar s$ contribute at tree-level, whereas the 
operators involving the third generation contribute to these processes only at the 1-loop level. Naively,
the former contributions are expected to dominate, but we will give a counterexample in the next subsection.
The experimental limits on the corresponding LFV decays are shown in Tab.~\ref{tab:lfvexp}.

\bigskip

\begin{figure}[htb]
        \begin{center}
        \includegraphics[width=0.45\textwidth]{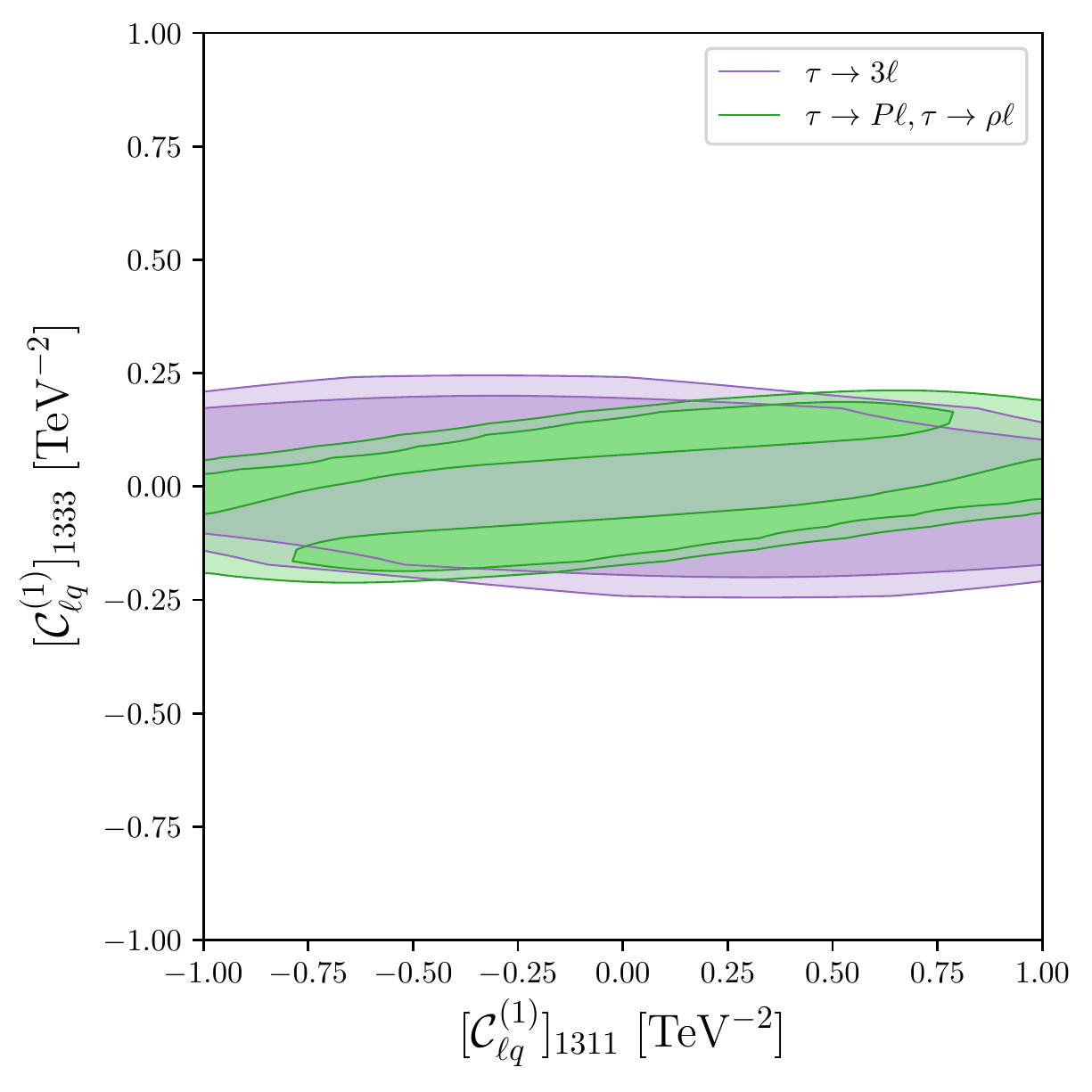}
        \includegraphics[width=0.45\textwidth]{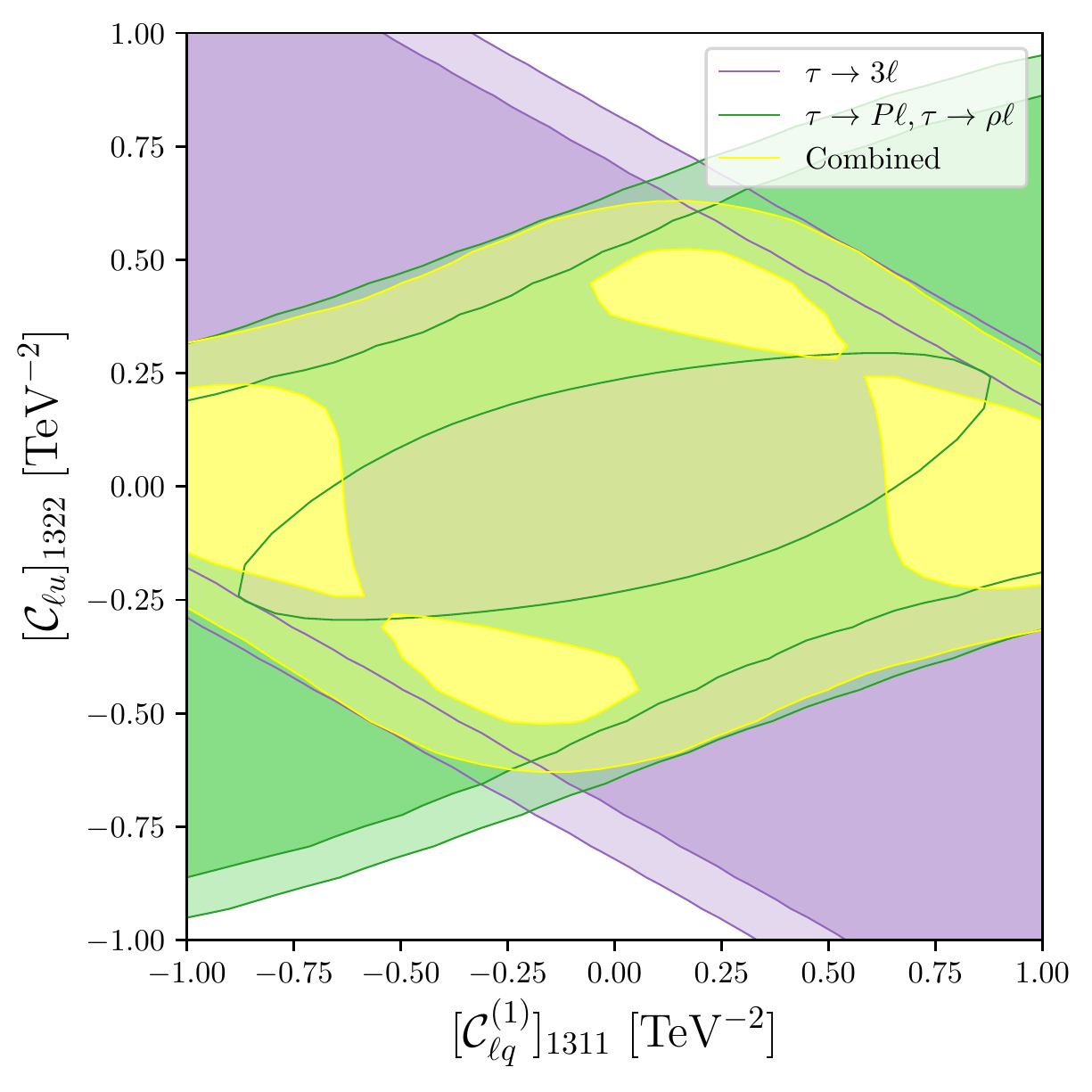}
        \end{center}
           \caption{\small The correlations between various LFV observables in the
        $\wc[(1)]{\ell q}{1311}$ - $\wc[(1)]{\ell q}{1333}$~ plane (left) and in
        $\wc[(1)]{\ell q}{1311}$ - $\wc[ ]{\ell u}{1322}$~ plane (right) .}
        \label{fig:corr2}
\end{figure}

\begin{table}[!htb]
  \begin{center}
\begin{tabular}{|lr|lr|}
\toprule
Observable &  Experimental value & Observable & Experimental value \\ \hline
$\mathcal{B}(Z\to e^\pm \mu^\mp)$ & $ <7.5 \times 10^{-7} $\cite{ATLAS:2014vur} & $\mathcal{B}(Z\to e^\pm \tau^\mp)$ & $<5.0 \times 10^{-6}$ \cite{ATLAS:2021bdj} \\
$\mathcal{B}(Z\to \mu^\pm \tau^\mp)$ & $< 6.5 \times 10^{-6} $ \cite{ATLAS:2021bdj} &  &\\ \hline
$\mathcal{B}(\tau^- \to e^-e^+e^- )$ & $<2.7\times 10^{-8}$ \cite{Hayasaka:2010np} &  $\mathcal{B}(\tau^- \to \mu^-\mu^+\mu^- )$ & $<2.1\times 10^{-8}$ \cite{Hayasaka:2010np}\\
$\mathcal{B}(\tau^- \to e^-\mu^+\mu^- )$ & $<2.7\times 10^{-8}$ \cite{Hayasaka:2010np} &  $\mathcal{B}(\tau^- \to \mu^- e^+e^- )$ & $<1.8\times 10^{-8}$ \cite{Hayasaka:2010np}\\
$\mathcal{B}(\tau^- \to \mu^- e^+ \mu^- )$ & $<1.7\times 10^{-8}$ \cite{Hayasaka:2010np} &  $\mathcal{B}(\tau^- \to  e^-\mu^+ e^- )$ & $<1.5\times 10^{-8}$ \cite{Hayasaka:2010np}\\ \hline
$\mathcal{B}(\mu^- \to e^-e^+e^-)$ & $<1.0\times 10^{-12}$ \cite{SINDRUM:1987nra} &  &   \\ \hline
$\mathcal{B}(\tau \to \phi \mu )$ & $<8.4\times 10^{-8}$ \cite{ParticleDataGroup:2016lqr} &  $\mathcal{B}(\tau \to  \phi e )$ & $<3.1\times 10^{-8}$ \cite{ParticleDataGroup:2016lqr}\\
$\mathcal{B}(\tau \to \rho \mu )$ & $<1.2\times 10^{-8}$ \cite{ParticleDataGroup:2016lqr} &  $\mathcal{B}(\tau \to  \rho e )$ & $<1.8\times 10^{-8}$ \cite{ParticleDataGroup:2016lqr} \\
$\mathcal{B}(\tau \to \pi \mu )$ & $< 1.1 \times 10^{-7}$ \cite{ParticleDataGroup:2016lqr} &  $\mathcal{B}(\tau \to  \pi e )$ & $<8.8\times 10^{-8}$ \cite{ParticleDataGroup:2016lqr}\\
\hline
\end{tabular}
\caption{\small The experimental upper limits on the LFV decays of $Z$-boson, $\tau$ and $\mu$ leptons.
 Note only the strongest limits are shown and the upper bounds correspond to $90\%$ CL for the $Z$ decays and 
at $90\%$ CL for all other decays.}
\label{tab:lfvexp}
\end{center}
\end{table}

\bigskip

\subsubsection{Correlations}

Since, the semileptonic operators can contribute to the LFV processes at tree-level as well 
as at the 1-loop, it would be interesting to see the relative importance of the loop-level vs. 
tree-level LFV effects. 
For example, the Wilson coefficient $\wc[(1)]{\ell q}{1311}$ could in principle give a tree level contribution 
to $\tau \to P e$ and $\tau \to \rho e$ processes. However, due to $SU(2)_L$ invariance,  
 the WET Wilson coefficients with $u \bar u$ and $d \bar d$ flavours get equal contributions 
on matching with $\wc[(1)]{\ell q}{1311}$. 
As a result, its net effect zero, because the $u\bar u$ and $d\bar d$ ~
Wilson coefficients enter with opposite sign in the branching ratios 
(see e.g. Eqs.~ (55) and (61) of Ref.~\cite{Aebischer:2018iyb}). 
However, $\wc[(1)]{\ell q}{1311}$ can still contribute at the 1-loop level through the EW corrections.

On the other hand, the Wilson coefficient $\wc[(1)]{\ell q}{1333}$ can not give tree-level contribution to 
 these LFV processes, but it can contribute at 1-loop to $\tau \to 3e$, $\tau \to e \mu^+ \mu^-$ as well as $\tau \to \rho e$ 
and $\tau \to P e$ processes. Interestingly, this loop induced effect on the $u\bar u$ and $d \bar d$~ WET coefficients 
is not equal but depends on the SM couplings $g_L^{d, {\rm SM}}$ and $g_L^{u, {\rm SM}}$ and hence it is not canceled in the 
branching ratios of our interest (see Eqs.~\eqref{eq:VLLedijkk} and \eqref{eq:VLLeuijkk}). 
In addition to VLL, the VLR (again with unequal sizes for the $u\bar u$ and $d \bar d$ coefficients) WET operators are 
also generated at the 1-loop level. 
In Fig.~\ref{fig:corr2} (left), the loose constraints on $\wc[(1)]{\ell q}{1311}$ as compared to 
$\wc[(1)]{\ell q}{1333}$ confirm these findings. On the right panel of the same figure, we show the impact of
 LFV constraints in the plane of $\wc[(1)]{\ell q}{1311}$ and $\wc[ ]{\ell u}{1322}$ Wilson coefficients.
In this case, since both operators contribute at the 1-loop level, they are found to be constrained at the 
similar level.

{\boldmath
\section{Sensitivities to the NP scale $\Lambda$}
\label{sec:scales}
}
\noindent
On the basis of our discussion about the RG running and the various low energy and the EW scale observables 
 identified in the previous sections, now we are in position to look at the 
highest possible scales for each operator that can 
be probed using these observables. In Eq.~\eqref{eq:1}, the Wilson coefficients are defined to be dimensionful quantities. 
However, these can be written in terms of the dimensionless parameters ${[c_X]}_{ijkl}$ defined by

\begin{equation}
\label{eq:smallc}
\wc[]{X}{ijkl} = \frac{[c_X]_{ijkl}}{\Lambda^2}.
\end{equation}
Assuming the presence of a single operator at the scale $\Lambda$, we perform combined fits for 
 each dimensionless parameter $[c_X]_{ijkl}$, using all measurements relevant for a given operator.
 From this we obtain the quantity ${\Lambda}/{\sqrt{{[c_X]}_{ijkl}}}$ using the central values 
of $[c_X]_{ijkl}$ from the fits.
This gives us a rough estimate of the scales that can be probed for each operator.

\begin{figure}[H]
        \begin{center}
        \includegraphics[width=\textwidth]{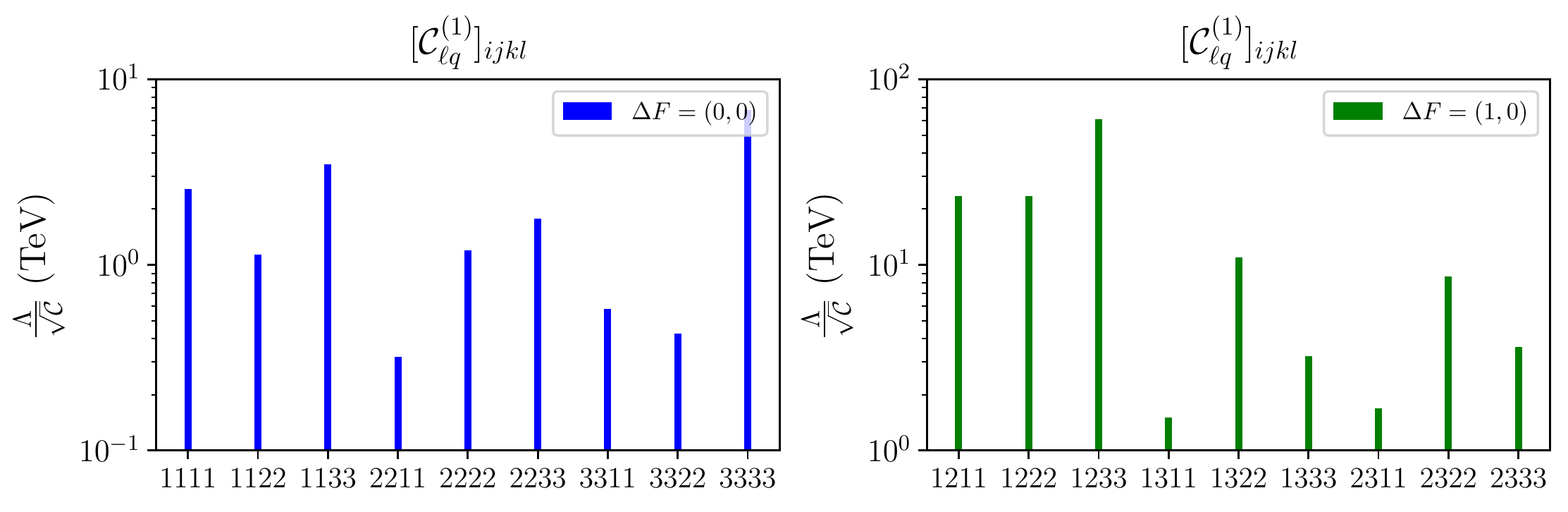}
        \end{center}
        \caption{\small The figure shows the sensitivity of semileptonic operators to the high scale $\Lambda$ normalized
 with dimensionless parameters ${[c_{\ell q}^{(1)}]}_{ijkl}$ as defined in Eq.~\eqref{eq:smallc}.}
        \label{fig:lq1}
\end{figure}
\begin{figure}[H]
        \begin{center}
        \includegraphics[width=\textwidth]{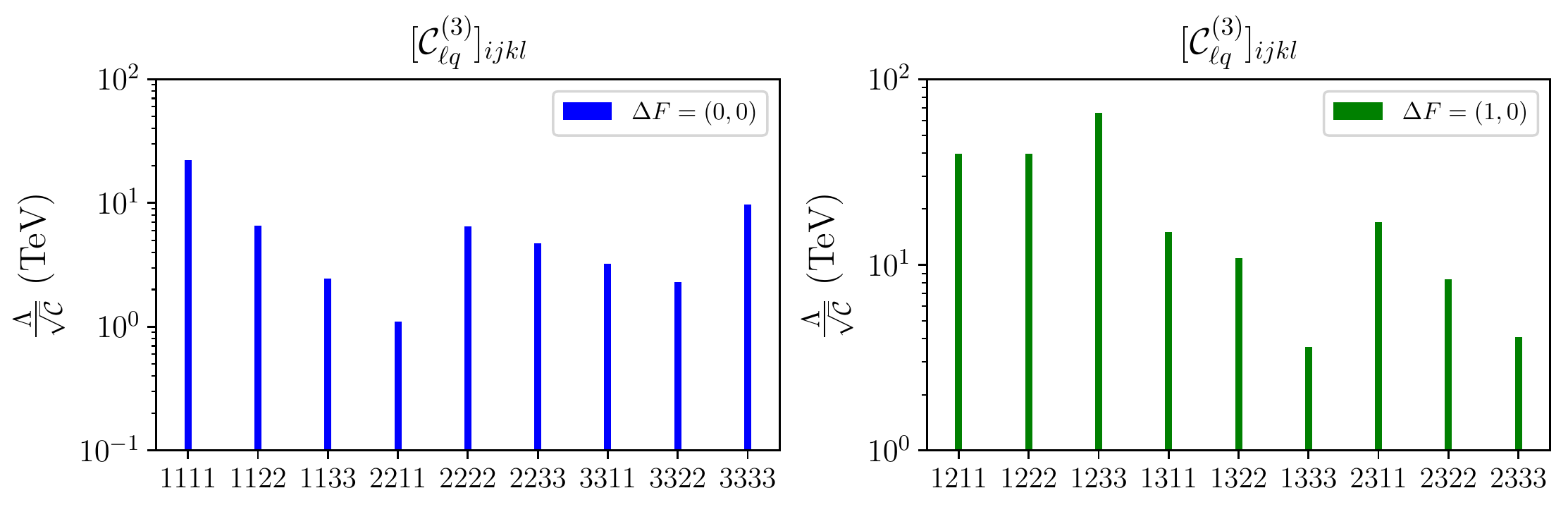}
        \end{center}
        \caption{\small Same as in Fig.~\ref{fig:lq1} except for ${[c_{\ell q}^{(3)}]}_{ijkl}$. }
\end{figure}

\begin{figure}[htb]
        \begin{center}
        \includegraphics[width=\textwidth]{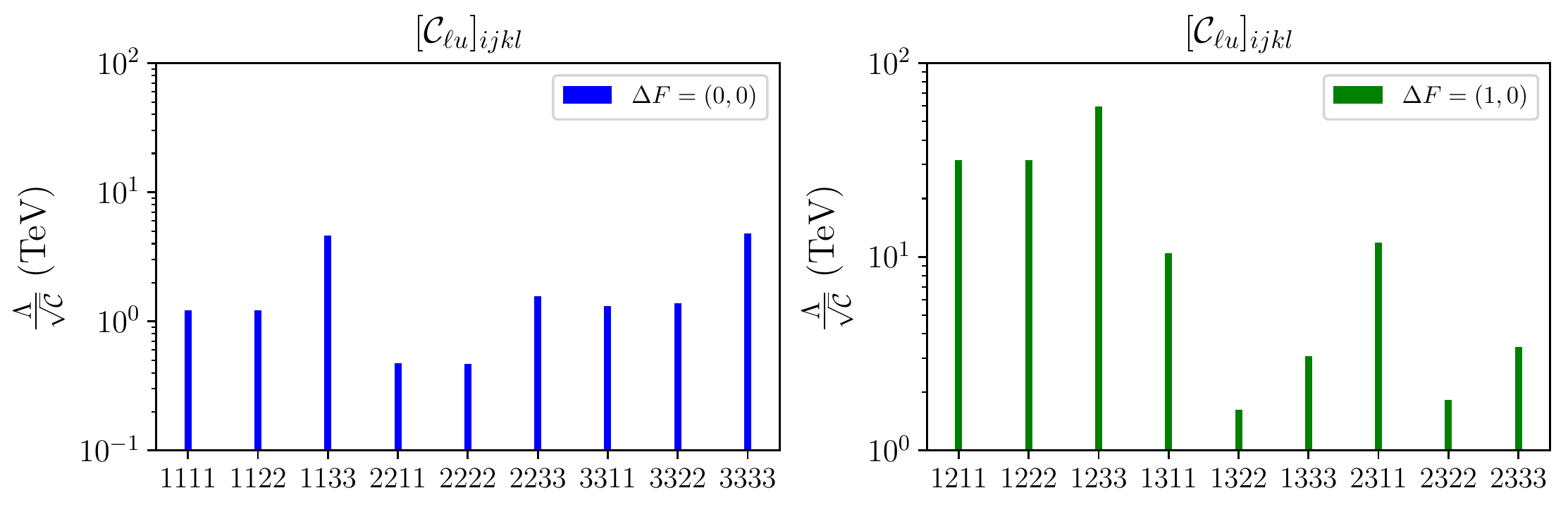}
        \end{center}
         \caption{\small Same as in Fig.~\ref{fig:lq1} except for ${[c_{\ell u}]}_{ijkl}$. }
\end{figure}

\begin{figure}[htb]
        \begin{center}
        \includegraphics[width=\textwidth]{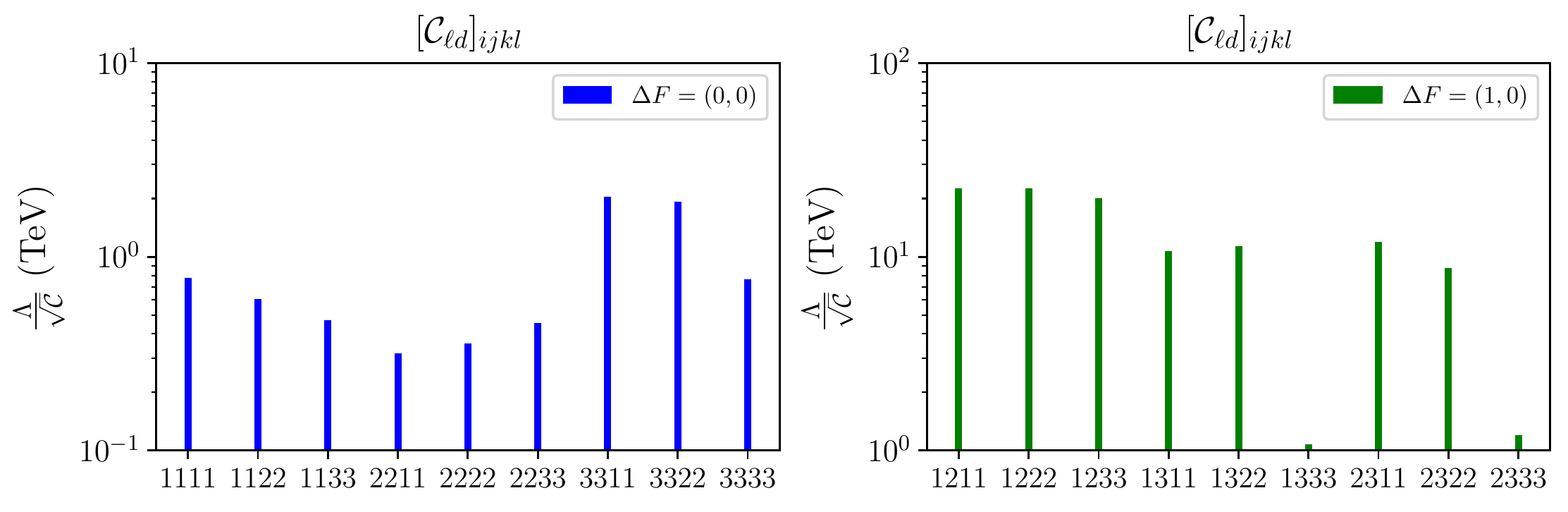}
        \end{center}
        \caption{\small Same as in Fig.~\ref{fig:lq1} except for ${[c_{\ell d}]}_{ijkl}$. }
\end{figure}

\begin{figure}[htb]
        \begin{center}
        \includegraphics[width=\textwidth]{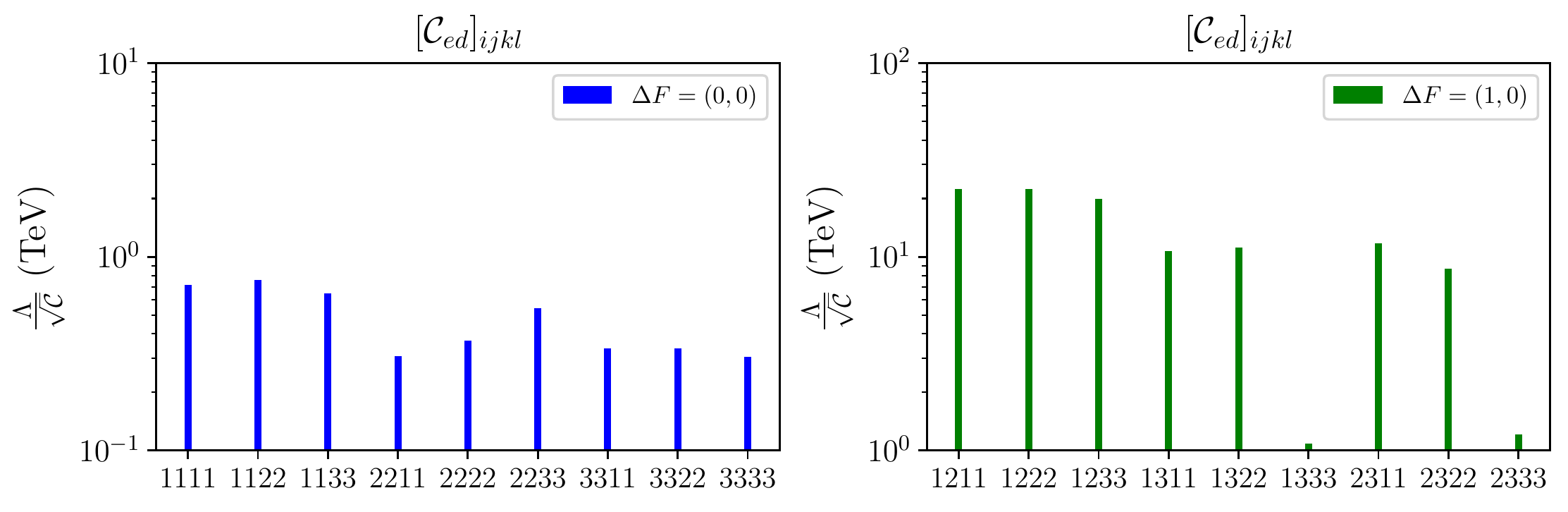}
        \end{center}
  \caption{\small Same as in Fig.~\ref{fig:lq1} except for ${[c_{ed}]}_{ijkl}$. }
\end{figure}
\begin{figure}[H]
        \begin{center}
        \includegraphics[width=\textwidth]{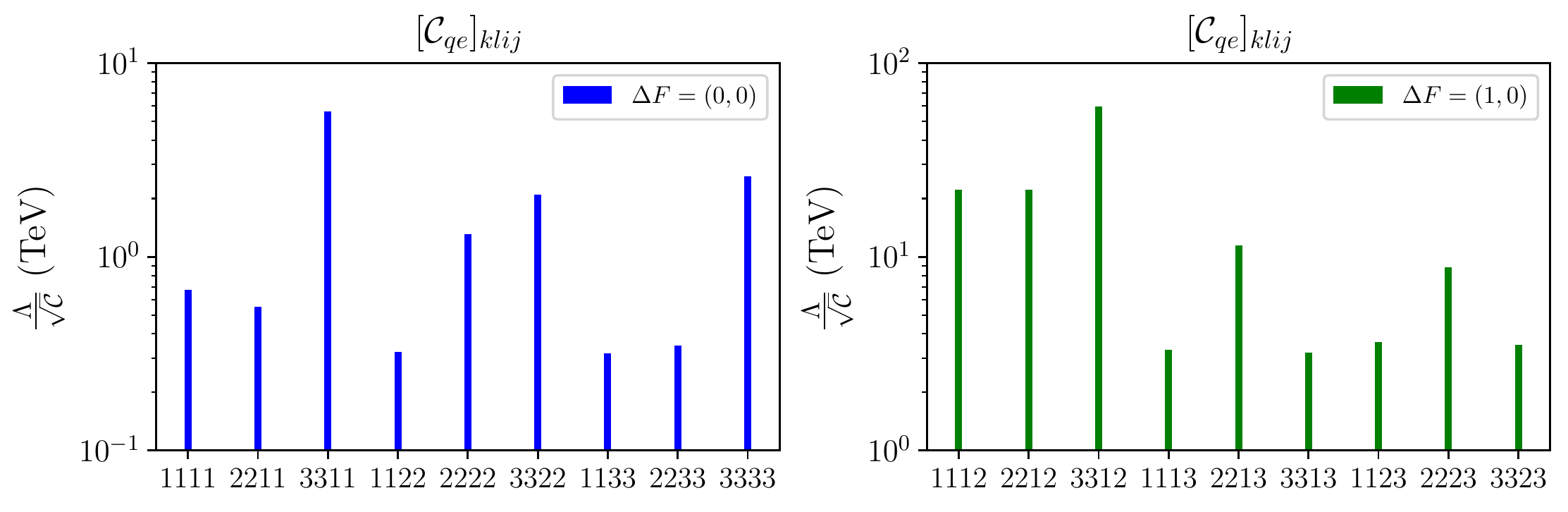}
        \end{center}
      \caption{\small Same as in Fig.~\ref{fig:lq1} except for ${[c_{qe}]}_{ijkl}$. }
       \label{fig:qe}
\end{figure}

 Note that in a given NP model, 
more than one operators can be simultaneously present which could change this simplified picture of 
single operator dominance. However, the goal of present work is to systematically analyze the low energy implications of 
each operator separately \emph{i.e}, to identify the most sensitive observables and assess their 
potential to constrain the individual operators. 
In  Figs.~\ref{fig:lq1}-\ref{fig:qe}, we show the central values for the lower bounds 
on the quantity ${\Lambda}/{\sqrt{{[c_X]}_{ijkl}}}$ 
(expressed in the units of TeV) for various Wilson coefficients under consideration. 
In the left and right panels, the results for  $\Delta F=(0,0)$ and $\Delta F=(1,0)$ operators 
are shown, respectively. Based on these results, we can now make following important observations:

\begin{figure}[htb]
        \begin{center}
        \includegraphics[width=\textwidth]{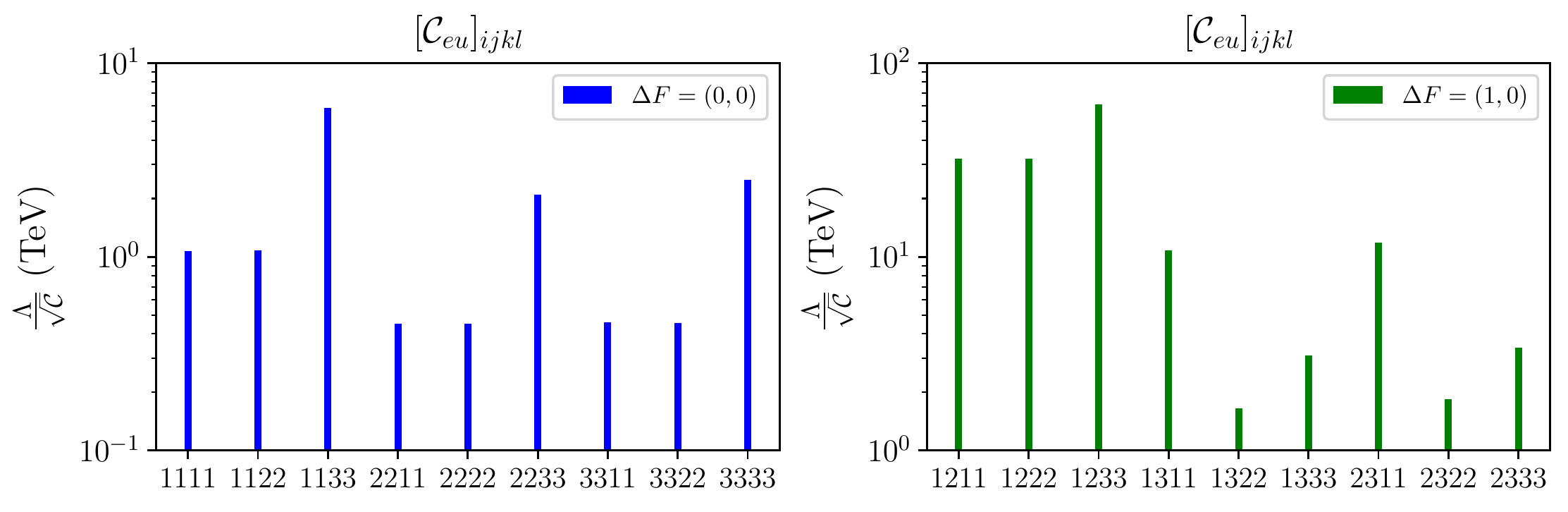}
        \end{center}
        \caption{\small Same as in Fig.~\ref{fig:lq1} except for ${[c_{e u}]}_{ijkl}$. }
\end{figure}

\begin{itemize}
\item The lower bounds on $\Delta F=(0,0)$ operators vary between $\mathcal{O}(\rm TeV)$ to $\mathcal{O}(10 \rm TeV)$, 
whereas the lower bounds on the $\Delta F=(1,0)$ operators go all the way up to $\mathcal{O}(100 \rm TeV)$, much beyond 
the present reach of the LHC.
\item Due to more stringent experimental limits on the branching ratio for the process $\mu \to 3e $ as compared to 
the $\tau$ LFV modes, the $\Delta F=(1,0)$ operators involving the $12jj$ flavour indices are very strongly 
constrained. As shown, the current lower bounds on such operators are always above the ballpark of 10 TeV.
\item Among $12jj$ operators, the ones having $jj=33$ are more strongly constrained as compared the operators 
with $jj=11$ or $22$. This can be attributed to the fact that the former operators can mix strongly with the 
 $\Delta F=(1,0)$ operators $\wc[(1)]{\phi \ell}{12}$, $\wc[(3)]{\phi \ell}{12}$, and $\wc[]{\phi e}{12}$ through top-Yukawa 
 interactions. In this regard, the numerical impact of the top-Yukawa is shown 
in Tab.~\ref{tab:wetleptonic} and Eq.~\eqref{eq:rge-df00-df10-yuk}. 
Clearly, the Wilson coefficients $\wc[]{\ell d}{ij33}$ and 
$\wc[]{ed}{1233}$ are exceptions here, because they do not mix with the contributing operators through top-Yukawa interactions.
\item Since the operators with the indices $13jj$, $23jj$ contribute even at tree-level to the LFV 
processes such as $\tau \to P\ell$ and $\tau \to \rho \ell$  for $jj=11$ or $22$, as a result these 
operators are in general more 
strongly constrained as compared to operators with $1333$ or $2333$ flavour indices. This is due to the reason that
the latter operators contribute only at the 1-loop to such 
processes or to purely leptonic LFV modes. However, in certain cases like 
$\wc[(1)]{\ell q}{1333}$, $\wc[(1)]{\ell q}{2333}$, 
 $\wc[]{\ell u}{1333}$ and $\wc[]{\ell u}{2333}$, the 1-loop effects due to the top-Yukawa can dominate over the tree-level effects.  
\item Except for $\wc[]{\ell d}{}$ and $\wc[]{ed}{}$,  
the $\Delta F=(0,0)$ operators involving $ii33$ for $ii=11,22$ or $33$ have the strongest bounds, because they contribute to the 
$Z$ and $W$ boson couplings through the top-Yukawa. 
\item Finally, among $\Delta F=(0,0)$ operators, $\wc[]{\ell d}{iijj}$ and $\wc[]{ed}{iijj}$ are found to 
be loosely constrained. In most cases, the lower bound on NP scale lies 
around $\mathcal{O}(1{\rm TeV})$ or below. This is again due to the fact that 
these operators do not exhibit the operator mixing due to large top-Yukawa coupling.
\end{itemize}

In \refapp{app:boundswcs}, we also provide the best-fit values along with $1\sigma$ errors 
for the dimensionless Wilson coefficients.

\section{Conclusions and Outlook}
\label{sec:conclusion}
\noindent
The SMEFT provides a convenient framework for parameterizing the NP effects beyond the SM.
 It is well known that the semileptonic operators which violate the quark flavour lead to 
 effects in the flavour violating decays of $B$ and $K$ mesons at low energy, which give rise to stringent 
constraints. In the view of current anomalies in the $B$-decays, the semileptonic operators  are 
of great interest in general.   However, it is important to 
probe the generic flavour structure of such operators. 
 In particular, often the quark and lepton flavour conserving operators and the ones which 
 violate only the lepton flavour are also generated in the NP models. To probe such operators it is important to know 
the type of observables to which these operators contribute.

In the present paper, we address this issue. We identify the low energy and the 
EW scale observables which can be used 
to probe a generic flavour 
structure of the semileptonic operators. However, in order to correctly predict the low energy behaviour of these operators, it 
is necessary to know the operator mixing pattern due to running from NP scale to the EW scale and then below. 
 To this end, by scrutinizing the ADMs due to the electroweak gauge as well as the Yukawas interactions 
one can find that such operators can mix with purely leptonic and $\psi^2 \phi^2 D$-type operators in the SMEFT. 
The former operators can contribute to the LFV decays of leptons, whereas the latter ones 
in addition also give corrections to the gauge boson couplings at the EW scale. 

Therefore, first we have identified the phenomenologically relevant terms in the ADMs and then 
 taking into account the WET and SMEFT RG running effects, we identified a list of observables which can be 
used to constrain the semileptonic operators having a generic flavour structure. 
We show that, through SMEFT RG running effects at the 1-loop level, the semileptonic operators 
can contribute to a variety of observables such as EWP observables, flavour violating
decays of $B$ and $K$-mesons $-$ involving neutral as well as charged current transitions, the 
LFV decays of leptons as well as the LFV $Z$-boson decays.
The main findings of the present study can be summarized as follows:

\newpage
The semileptonic operators with flavour indices $iijj$ for $ii,jj=11,22$ or $33$ contribute to the 
EWP observables at the 1-loop level. We have identified two different types of contributions in this context (1) the 
operator mixing with $\psi^2 \phi^2 D$ operators through gauge interactions affect the 
$W$ and $Z$ boson vertices at the EW scale. 
Then, through the top-Yukawa interactions, there are additional contributions from 
the operators involving third generation in the quark current, \emph{i.e.},  
$\wc[(1)]{\ell q}{ii33}$, $\wc[(3)]{\ell q}{ii33}$, $\wc[]{\ell u}{ii33}$, 
$\wc[]{eu}{ii33}$ and $\wc[]{qe}{33ii}$. (2) The $W$ and $Z$ boson vertices also receive 
corrections from $\wc[(3)]{\ell q}{1133}$ and $\wc[(3)]{\ell q}{2233}$ operators 
via shifts in the $\delta g_Z$ and $\delta \sin^2 \theta_W$. We present the relative impact of the 
running due to the gauge and top-Yukawas on the $W/Z$ couplings at the EW scale due to 
semileptonic operators at $\Lambda$. 
%
In addition, the quark flavour conserving operators such as, $\wc[(1)]{\ell q}{ijkl}$, $\wc[]{ed}{ijkl}$, 
$\wc[]{\ell d}{ijkl}$, and $\wc[]{q e}{ijkl}$ with $ijkl=1122,1133,2222,2233$ can be constrained 
by $b\to s \ell^+ \ell^-$ processes through the back-rotation effect.
%
We have shown that the Wilson coefficients $\wc[(1)]{\ell q}{ijkl}$ with $ijkl=1122$, $3322$, 
$3311$, or $1111$ can contribute to 
various charged current processes via operator mixing with $\wc[(3)]{\ell q}{ijkl}$. This effects goes through 
the EW corrections.
%

The semileptonic operators which violate the lepton flavour, while conserving the 
quark flavour can contribute to the 
LFV decays of $\tau$ and $\mu$ leptons. Some of the semileptonic operators contribute at tree-level 
 to $\tau \to P\ell$ and $\tau \to \rho \ell$ processes. On the other hand, purely leptonic LFV decays 
such as $\tau \to 3\mu$ and $\mu \to 3 e$ etc., 
 are generated only at the 1-loop level. Again this happens through operator mixing which depends on the
 gauge as well as Yukawa interactions. In this regard, we find that depending upon the flavour 
structures the Yukawas play an important in constraining such 
operators. We have also studied the relative impact of RG running due to various sources  on the contributing 
WET operators at the low scale. These results are presented in Tab.~\ref{tab:wetleptonic}~ for purely leptonic 
WET operators and in Tabs.~\ref{tab:wethadronic}~ for semileptonic WET operators. 
%

Finally, using the latest measurements, we derived lower bounds on the cut-off scale $\Lambda$ for each semileptonic 
operator under consideration.
We observe that depending upon the flavour structure, the RG induced constraints can lead to 
  sensitivities to very high NP scales varying between $\mathcal{O} ({\rm 1 TeV})$ to $\mathcal{O}(100{\rm TeV})$.
Finally, for a fixed value of $\Lambda =3$ TeV, we also provide allowed ranges for the semileptonic 
Wilson coefficients.

\section*{Note Added}
While finalizing our paper we notice Ref.~\cite{Calibbi:2021pyh} on arXiv which presented a study of the LFV decays through $Z$-flavour violation in the context of future 
colliders.   The discussion on the LFV decays partly overlaps with the present work.
However, the aim of the current study is not limited to the LFV modes but is to analyze both flavour conserving as well as the 
flavour violating processes to which semileptonic operators can contribute at tree-level or through operator mixing.

\section*{Acknowledgments}
J. K. thanks Andrzej Buras and Jason Aebischer for carefully reading the manuscript and for the useful comments. 
J.K. is financially supported by the Alexander von Humboldt Foundation's postdoctoral research 
fellowship.

\appendix

{\boldmath
\section{Tree-level shifts in the dim-4 gauge boson couplings}
\label{app:shiftsew}
}

For completeness, in this section we collect formulae for the tree-level shifts 
due to SMEFT in the dim-4 $Z$ and $W$ boson  
couplings at the EW scale. For this purpose we closely follow Ref.~\cite{Brivio:2017vri}. 
The shifts in the $Z$ boson couplings $\delta(g_X^Y)_{ij}^{\rm dir}$ 
(see Eq.~\eqref{eq:def-delta} for definition) get tree-level contributions due to 
$\psi^2 \phi^2 D$-type operators in the SMEFT. At the EW scale for the quarks these are given by 
\begin{eqnarray}
\label{eq:zcouplingsa}
\delta (g_L^u)_{ij}^{\rm dir} &=& -\frac{v^2}{2} V_{im}  \left ( \wc[(1)]{\phi q}{mn}  - \wc[(3)]{\phi q}{mn}\right ) V^\dagger_{nj}\,,\\ 
\quad \delta (g_R^u)_{ij}^{\rm dir} &=& -\frac{ v^2}{2}  \wc[]{\phi u}{ij}\,,   \\
\delta (g_L^d)_{ij}^{\rm dir} & = & -\frac{v^2}{2} \left ( \wc[(1)]{\phi q}{ij}  + \wc[(3)]{\phi q}{ij}\right )\,,  \\ 
\quad \delta (g_R^d)_{ij}^{\rm dir} &=& -\frac{ v^2}{2}  \wc[]{\phi d}{ij}.
\end{eqnarray}
Similarly, for the leptons we have
\begin{eqnarray}   
\delta (g_L^\nu)_{ij}^{\rm dir} &=& -\frac{v^2}{2} \left ( \wc[(1)]{\phi \ell}{ij}  - \wc[(3)]{\phi \ell}{ij}\right )\,,   \\
\delta (g_L^e)_{ij}^{\rm dir} & = & -\frac{v^2}{2} \left ( \wc[(1)]{\phi \ell}{ij}  + \wc[(3)]{\phi \ell}{ij}\right )\,,\\ 
\quad \delta (g_R^e)_{ij}^{\rm dir } &=& -\frac{v^2}{2}  \wc[]{\phi e}{ij}.
\label{eq:zcouplingsz}
\end{eqnarray}
The CKM elements appearing in the expressions for 
 $\delta (g_L^u)_{ij}^{\rm dir}$ are needed due to our 
Warsaw-down basis choice for the SMEFT operators. 
In addition, the shifts in the parameters $g_Z$ and $\sin \theta_W$ are 
given by
\begin{equation} 
\label{eq:gztree}
\delta g_Z=  - \frac{v^2}{2} \left (  \wc[(3)]{\phi \ell}{11} +\wc[(3)]{\phi \ell }{22}   - 
\frac{\wc[]{\ell \ell}{1221}}{ 2}   \right ) \,,
\end{equation}
 and 
\begin{equation}
\label{eq:s2wtree} 
\delta \sin^2{\theta_W} = \frac{ v^2 \sin^2{2 \theta_W}}{4  \cos{2 \theta_W}} \left (\wc[(3)]{\phi \ell}{11}+ \wc[(3)]{\phi \ell}{22} 
- \frac{ \wc[]{\ell \ell}{1221} }{2} \right ). 
\end{equation}
The $W$ boson couplings can be analogously parameterized as 
\begin{eqnarray}
\delta (\varepsilon_L^\ell)_{ij}^{\rm dir} &=& v^2 \wc[(3)]{\phi \ell}{ij} \,, \\
\delta (\varepsilon_L^q)_{ij}^{\rm dir} &=& v^2 V_{im}  \wc[(3)]{\phi q}{mj} .
\end{eqnarray}

Since, we are interested to study the effects of only the semileptonic operators on the EWP observables, we have 
ignored additional corrections due to $\wc[]{\phi D}{}$ and $\wc[]{\phi W B}{}$ operators\cite{Brivio:2017vri}. 
In this regard, we have checked that these operators can not be generated from semileptonic operators 
 via operator mixing.

{\boldmath
\section{RG induced shifts in the dim-4 and dim-6 operators}
\label{app:comp-gauge-yuk}
}
\noindent
Depending upon the scale and the interactions involved, there are three types of RGEs, 
\emph{i.e.}, due to the gauge and Yukawa interactions in the SMEFT 
and due to QCD+QED interactions in the WET. 
In this section, we analyze the relative impact of three different types of RG runnings on the dim-4 
and dim-6 operators which contribute to the EWP observables and the LFV processes at the low energy.

\bigskip

\subsection{Dim-4 $Z$ boson couplings}

In Tab.~\ref{tab:rge-lq1-1133}-\ref{tab:rge-qe-3311}, we show 
the $Z$ and $W$ boson gauge couplings at the EW scale for a given semileptonic 
operator introduced at the high scale.

\begin{table}[H]
\begin{center}
\begin{adjustbox}{width=0.6\textwidth}\begin{tabular}{|rrr|}
\toprule
 ${\rm Coupling}(m_Z)$ & Gauge Couplings & Yukawa Couplings \\ \hline
\multicolumn{1}{|c}{$\wc[(1)]{\ell q}{ii33}(\Lambda) =1$} & &\\
 $\delta (g_L^u)_{11}$ &               - &            2.28 \\
 $\delta (g_L^u)_{22}$ &               - &            2.28 \\
 $\delta (g_L^d)_{11}$ &               - &             -1.68 \\
 $\delta (g_L^d)_{22}$ &               - &             -1.68 \\
 $\delta (g_L^d)_{33}$ &            -4.02 &             -5.67 \\
 $\delta (g_L^e)_{ii}$ &           3.58 &          220.63 \\
 $\delta (g_L^e)_{jj}$ &               - &             -2.87 \\
 $\delta (g_R^u)_{11}$ &               - &            1.19 \\
 $\delta (g_R^u)_{22}$ &               - &            1.19 \\
 $\delta (g_R^e)_{kk}$ &               - &             -1.79 \\
\bottomrule
\end{tabular}
\end{adjustbox}                                            

\end{center}
\caption{\small The shifts in the $Z$ boson couplings with fermions \emph{i.e.,} $\delta (g_X^Y)_{ii} \times 10^5 $ 
at $\mu \simeq 91$ GeV due to semileptonic Wilson coefficient $\wc[(1)]{\ell q}{ii33}(\Lambda) =1 {\rm TeV^{-2}}$ at $\Lambda =1 {\rm TeV}$.
 Here $ii=11$ or $22$, $jj =11,22,33 \ne ii$ and  $kk=11,22,33$. In the second and third column, the RG running due to gauge interactions only and 
gauge + Yukawa interactions is shown, respectively. Note, only the non-zero entries are shown.}
\label{tab:rge-lq1-1133}
\end{table}
\begin{table}[H]
\begin{center}
\begin{adjustbox}{width=0.6\textwidth}\begin{tabular}{|rrr|}
\toprule
 ${\rm Coupling}(m_Z)$ & Gauge Couplings & Yukawa Couplings \\ \hline
\multicolumn{1}{|c}{$\wc[(1)]{\ell q}{3333}(\Lambda) =1$} & &\\
 $\delta (g_L^d)_{33}$ &            -4.04 &             -4.27 \\
 $\delta (g_L^e)_{33}$ &           3.54 &          222.69 \\
\bottomrule
\end{tabular}
\end{adjustbox}                                            

\end{center}
\caption{Same as in Tab.~\ref{tab:rge-lq1-1133} except for $\wc[(1)]{\ell q}{3333}(\Lambda)$.}
\label{tab:rge-lq1-3333}
\end{table}
\begin{table}[H]
\begin{center}
\begin{adjustbox}{width=0.6\textwidth}\begin{tabular}{|rrr|}
\toprule
 ${\rm Coupling}(m_Z)$ & Gauge Couplings & Yukawa Couplings \\
\hline
\multicolumn{1}{|c}{$\wc[(3)]{\ell q}{ii33}(\Lambda) =1$} & &\\
 $\delta (g_L^u)_{11}$ &               - &           -181.11 \\
 $\delta (g_L^u)_{22}$ &               - &           -181.11 \\
 $\delta (g_L^d)_{11}$ &               - &          147.73 \\
 $\delta (g_L^d)_{22}$ &               - &          147.73 \\
 $\delta (g_L^d)_{33}$ &          12.86 &          160.61 \\
 $\delta (g_L^e)_{ii}$ &          38.66 &           17.72 \\
 $\delta (g_L^e)_{jj}$ &               - &          214.52 \\
%
 $\delta (g_R^u)_{11}$ &               - &            -66.79 \\
 $\delta (g_R^u)_{22}$ &               - &            -66.79 \\
 $\delta (g_R^d)_{kk}$ &               - &           33.40 \\
%
$\delta (g_R^e)_{kk}$ &               - &          100.19 \\
\bottomrule
\end{tabular}
\end{adjustbox}                                            

\end{center}
\caption{Same as in Tab.~\ref{tab:rge-lq1-1133} except for $\wc[(3)]{\ell q}{ii33}(\Lambda)$ with $ii=11$ or $22$.}
\label{tab:rge-lq3-1133}
\end{table}
\begin{table}[H]
\begin{center}
\begin{adjustbox}{width=0.6\textwidth}\begin{tabular}{|rrr|}
\toprule
 ${\rm Coupling}(m_Z)$ & Gauge Couplings & Yukawa Couplings \\ \hline
\multicolumn{1}{|c}{$\wc[(3)]{\ell q}{3333}(\Lambda) =1$} & &\\
 $\delta (g_L^d)_{33}$ &          12.93 &           12.61 \\
 $\delta (g_L^e)_{33}$ &          38.76 &           -197.61 \\
\bottomrule
\end{tabular}
\end{adjustbox}                                            

\end{center}
\caption{Same as in Tab.~\ref{tab:rge-lq1-1133} except for $\wc[(3)]{\ell q}{3333}(\Lambda)$.}
\label{tab:rge-lq3-3333}
\end{table}
\begin{table}[H]
\begin{center}
\begin{adjustbox}{width=0.6\textwidth}\begin{tabular}{|rrr|}
\toprule
 ${\rm Coupling}(m_Z)$ & Gauge Couplings & Yukawa Couplings \\ \hline
\multicolumn{1}{|c}{$\wc[]{\ell u}{ii33}(\Lambda) =1$} & &\\
$\delta (g_L^e)_{ii}$ &           7.85 &           -208.36 \\
\bottomrule
\end{tabular}
\end{adjustbox}                                            

\end{center}
\caption{Same as in Tab.~\ref{tab:rge-lq1-1133} except for $\wc[]{\ell u}{ii33}(\Lambda)$ with $ii=11,22$ or $33$.}
\label{tab:rge-lu-1133}
\end{table}
\begin{table}[H]
\begin{center}
\begin{adjustbox}{width=0.6\textwidth}\begin{tabular}{|rrr|}
\toprule
 ${\rm Coupling}(m_Z)$ & Gauge Couplings & Yukawa Couplings \\
\hline
\multicolumn{1}{|c}{$\wc[]{\ell d}{ii33}(\Lambda) =1$} & &\\
 $\delta (g_L^e)_{ii}$ &            -3.95 &             -3.89 \\
 $\delta (g_R^d)_{33}$ &            -3.92 &             -3.92 \\
\bottomrule
\end{tabular}
\end{adjustbox}                                            

\end{center}
\caption{Same as in Tab.~\ref{tab:rge-lq1-1133} except for $\wc[]{\ell d}{ii33}(\Lambda)$ with $ii=11,22$ or $33$.}
\label{tab:rge-ld-1133}
\end{table}
\begin{table}[H]
\begin{center}
\begin{adjustbox}{width=0.6\textwidth}\begin{tabular}{|rrr|}
\toprule
 ${\rm Coupling}(m_Z)$ & Gauge Couplings & Yukawa Couplings \\ \hline
\multicolumn{1}{|c}{$\wc[]{eu}{ii33}(\Lambda) =1$} & &\\
 $\delta (g_R^e)_{ii}$ &           7.81 &           -211.10 \\
\bottomrule
\end{tabular}
\end{adjustbox}                                            

\end{center}
\caption{Same as in Tab.~\ref{tab:rge-lq1-1133} except for $\wc[]{eu}{ii33}(\Lambda)$ with $ii=11,22$ or $33$.}
\label{tab:rge-eu-1133}
\end{table}
\begin{table}[H]
\begin{center}
\begin{adjustbox}{width=0.6\textwidth}\begin{tabular}{|rrr|}
\toprule
 ${\rm Coupling}(m_Z)$ & Gauge Couplings & Yukawa Couplings \\ \hline
\multicolumn{1}{|c}{$\wc[(1)]{qe}{33ii}(\Lambda) =1$} & &\\
 $\delta (g_L^d)_{33}$ &            -3.91 &             -4.13 \\
 $\delta (g_R^e)_{ii}$ &           3.87 &          220.04 \\
\bottomrule
\end{tabular}
\end{adjustbox}                                            

\end{center}
\caption{Same as in Tab.~\ref{tab:rge-lq1-1133} except for $\wc[]{qe}{33ii}(\Lambda)$ with $ii=11,22$ or $33$.}
\label{tab:rge-qe-3311}
\end{table}


\subsection{Dim-4 $W$ boson couplings}

In Tab.~\ref{tab:rge-lq3-1133W} and \ref{tab:rge-lq3-3333W} we show impact of RG running due to 
Yukawas on the $W$ couplings to fermions at the EW scale.

\begin{table}[H]
\begin{center}
\begin{adjustbox}{width=0.6\textwidth}\begin{tabular}{|rrr|}
\toprule
              ${\rm Coupling}(m_W)$ & Gauge Couplings & Yukawa Couplings \\ \hline
\multicolumn{1}{|c}{$\wc[(3)]{\ell q}{ii33}(\Lambda) =1$} & &\\
    $\delta (\varepsilon_L^q)_{11}$ &               - &          -328.37 \\
    $\delta (\varepsilon_L^q)_{22}$ &               - &          -328.37 \\
 $\delta (\varepsilon_L^\ell)_{ii}$ &          -77.31 &            52.11 \\
 $\delta (\varepsilon_L^\ell)_{jj}$ &               - &          -328.37 \\
\bottomrule
\end{tabular}
\end{adjustbox}                                            

\end{center}
\caption{\small The shifts in the $W$ boson couplings \emph{i.e.,} $\delta (\varepsilon_L^Y)_{ii} \times 10^5 $
to fermions at $\mu \simeq 91$ GeV due to semileptonic
operator $\wc[(3)]{\ell q}{ii33}(\Lambda) =1 {\rm TeV^{-2}}$ at $\Lambda =1 {\rm TeV}$. 
 Here $ii=11$ or $22$, and $jj=11,22,33 \ne ii$. In the second and third column the RG running 
due to gauge only and
gauge + Yukawa interactions is included, respectively. Note, only the non-zero entries are shown.}
\label{tab:rge-lq3-1133W}
\end{table}
\begin{table}[H]
\begin{center}
\begin{adjustbox}{width=0.6\textwidth}\begin{tabular}{|rrr|}
\toprule
              ${\rm Coupling}(m_W)$ & Gauge Couplings & Yukawa Couplings \\
\hline
\multicolumn{1}{|c}{$\wc[(3)]{\ell q}{3333}(\Lambda) =1$} & &\\
 $\delta (\varepsilon_L^\ell)_{33}$ &          -77.46 &           381.44 \\
\bottomrule
\end{tabular}
\end{adjustbox}                                            

\end{center}
\caption{Same as in Tab.~\ref{tab:rge-lq3-1133W} except for $\wc[(3)]{\ell q}{3333}(\Lambda)$.}
\label{tab:rge-lq3-3333W}
\end{table}

\newpage

\subsection{Leptonic dim-6 WET operators}
In Tab.~\ref{tab:wetleptonic}, we present the numerical values for the leptonic WET Wilson coefficients by setting 
 the SMEFT Wilson coefficients to 1 ${\rm TeV^{-2}}$ at  $\Lambda=1 {\rm TeV}$. 
In order to study the roles played by different kinds of RG effects, we present 
three kind of numbers, (1) with only WET (QED+QCD) running, (2) WET + SMEFT running 
due to gauge interactions, and (3) the full WET+ SMEFT gauge + Yukawa running.
As we can see, except for $\wc[]{\ell d}{1333}$ and $\wc[]{ed}{1333}$, the top-Yukawa 
 effects always play an important role for the operators involving third generation quarks. 
\begin{table}[H]
\begin{center}
\begin{adjustbox}{width=0.49\textwidth}
\begin{tabular}{|rrrr|} \hline
\toprule
${\rm Coefficent}$ &         QCD+QED& Gauge Couplings& Yukawa Couplings\\ \hline
\multicolumn{1}{|c}{$[\mathcal{C}_{\ell q}^{(1)}]_{1311}(\Lambda) = 1 $} && & \\ 
         $[C_{ee}^{V,LL}]_{\ell\ell 13}$ &      3.41 &      5.48 &        5.48 \\
         $[C_{ee}^{V,LR}]_{13\ell\ell}$ &      3.41 &       5.64 &        5.63 \\
\hline


\multicolumn{1}{|c}{$[\mathcal{C}_{\ell q}^{(1)}]_{1333}(\Lambda) = 1 $} &&&\\ 
         $[C_{ee}^{V,LL}]_{\ell\ell 13}$ &      -3.40 &       -1.49 &        -35.25 \\
         $[C_{ee}^{V,LR}]_{13\ell\ell}$ &      -3.40 &       -1.33 &          36.98 \\
\hline

\multicolumn{1}{|c}{$[\mathcal{C}_{\ell q}^{(3)}]_{1311}(\Lambda) = 1 $} && & \\
         $[C_{ee}^{V,LL}]_{\ell\ell 13}$ &      -10.24 &       -15.91 &        15.91 \\
         $[C_{ee}^{V,LR}]_{13\ell\ell}$ &      -10.24 &       -16.66 &         16.66\\
\hline


\multicolumn{1}{|c}{$[\mathcal{C}_{\ell q}^{(3)}]_{1333}(\Lambda) = 1 $} &&&\\
         $[C_{ee}^{V,LL}]_{\ell\ell 13}$ &      -3.42 &       -8.42 &        47.68 \\
         $[C_{ee}^{V,LR}]_{13\ell\ell}$ &      -3.42 &       -9.17 &        -29.10 \\
\hline


\multicolumn{1}{|c}{$[\mathcal{C}_{\ell u}]_{1311}(\Lambda) = 1 $} &&&\\ 
         $[C_{ee}^{V,LL}]_{\ell\ell 13}$ &      6.83 &       10.37 &        10.37 \\
         $[C_{ee}^{V,LR}]_{13\ell\ell}$ &       6.83 &       10.67 &        10.67 \\
\hline


\multicolumn{1}{|c}{$[\mathcal{C}_{\ell u}]_{1333}(\Lambda) = 1 $} &&&\\
         $[C_{ee}^{V,LL}]_{\ell\ell 13}$ &      -          &       3.60 &        -39.64 \\
         $[C_{ee}^{V,LR}]_{13\ell\ell}$ &       -          &       3.90 &        -39.08 \\
\hline

\multicolumn{1}{|c}{$[\mathcal{C}_{\ell d}]_{1311}(\Lambda) = 1 $} &&&\\
         $[C_{ee}^{V,LL}]_{\ell\ell 13}$ &      -3.14          &       -5.24 &        5.24 \\
         $[C_{ee}^{V,LR}]_{13\ell\ell}$ &       -3.14          &       -5.39 &        5.39 \\
\hline

\multicolumn{1}{|c}{$[\mathcal{C}_{\ell d}]_{1333}(\Lambda) = 1 $} &&&\\
         $[C_{ee}^{V,LL}]_{\ell\ell 13}$ &      -3.14          &       -5.24 &        5.23 \\
         $[C_{ee}^{V,LR}]_{13\ell\ell}$ &       -3.14          &       -5.39 &        5.40 \\
\hline


\multicolumn{1}{|c}{$[\mathcal{C}_{e u}]_{1311}(\Lambda) = 1 $} &&&\\ 
         $[C_{ee}^{V,LR}]_{\ell\ell 13}$ &           6.83 &            10.88 &             -10.88 \\
         $[C_{ee}^{V,RR}]_{\ell\ell 13}$ &           6.83 &            10.88 &             -10.88 \\
\hline                          

\multicolumn{1}{|c}{$[\mathcal{C}_{e u}]_{1333}(\Lambda) = 1 $} &&&\\ 
         $[C_{ee}^{V,LR}]_{\ell\ell 13}$ &           - &            3.94 &             -35.65 \\
         $[C_{ee}^{V,RR}]_{\ell\ell 13}$ &           - &            3.94 &             -35.65 \\

\hline

\multicolumn{1}{|c}{$[\mathcal{C}_{e d}]_{1311}(\Lambda) = 1 $}&&& \\ 
         $[C_{ee}^{V,LR}]_{\ell\ell 13}$ &           -3.14 &            -5.33 &             5.33 \\
         $[C_{ee}^{V,RR}]_{\ell\ell 13}$ &           -3.14 &            -5.33 &             5.33 \\
\hline

\multicolumn{1}{|c}{$[\mathcal{C}_{e d}]_{1333}(\Lambda) = 1 $} &&&\\
         $[C_{ee}^{V,LR}]_{\ell\ell 13}$ &           -3.14 &            -5.33 &             -5.32 \\
         $[C_{ee}^{V,RR}]_{\ell\ell 13}$ &           -3.14 &            -5.33 &             -5.32 \\
\hline

\multicolumn{1}{|c}{$[\mathcal{C}_{qe}]_{1311}(\Lambda) = 1$}&&& \\ 
         $[C_{ee}^{V,LR}]_{\ell\ell 13}$ &           3.41 &            5.35 &             -5.35 \\
         $[C_{ee}^{V,RR}]_{\ell\ell 13}$ &           3.41 &            5.35 &             -5.35 \\
\hline

\multicolumn{1}{|c}{$[\mathcal{C}_{q e}]_{1333}(\Lambda) = 1 $} &&&\\ 
         $[C_{ee}^{V,LR}]_{\ell\ell 13}$ &           -3.4 &            -1.44 &             -37.66 \\
         $[C_{ee}^{V,RR}]_{\ell\ell 13}$ &           -3.400 &            -1.44 &             -37.66 \\

\hline
\end{tabular}
\end{adjustbox}

\caption{\small The impact of semileptonic operators on the low energy purely leptonic $\Delta F=(1,0)$ Wilson coefficients 
(in $10^{-9}~ {\rm TeV^{-2}}$ units) at $\mu_{\rm low}=2$GeV is shown. The second column refers 
to the values with only WET (QED+QCD) running, the third column refers to the WET + SMEFT running 
due to gauge interactions, and the fourth column refers to the full WET+ SMEFT 
gauge + Yukawa running. Here, the indices $\ell\ell=11$,$22$, or $33$ and $\Lambda=1$TeV. 
The SMEFT Wilson coefficients are set to 1 (in $\rm TeV^{-2}$ units) at $\Lambda$.   }
\label{tab:wetleptonic}
\end{center}
\end{table}

\subsection{Semileptonic dim-6 WET operators}
\label{app:rgesemilep}

In Tab.~\ref{tab:wethadronic}, we present the low energy semileptonic WET Wilson coefficients by setting 
various semileptonic SMEFT Wilson coefficients, involving the third generation of quarks, equal to 1 ${\rm TeV^{-2}}$ 
at $\Lambda=1 {\rm TeV}$. Again, in order to study the roles played by different 
kinds of RG running, we have presented three kind of numbers. As found in the case of WET leptonic operators, 
 except for the Wilson coefficients $\wc[]{\ell d}{1333}$ and $\wc[]{ed}{1333}$, the top-Yukawa always plays an
important role.

\begin{table}[H]
\begin{center}
\begin{adjustbox}{width=0.59\textwidth}
\begin{tabular}{|rrrr|} \hline
\toprule
${\rm Coefficent}$ &         QCD+QED& Gauge Couplings& Yukawa Couplings\\ \hline
\multicolumn{1}{|c}{$[\mathcal{C}_{\ell q}^{(1)}]_{1333}(\Lambda) = 1 $} && & \\ 
         $[C_{ed}^{V,LL}]_{13jj}$ &     -1.13 &      -0.47 &       -59.92 \\
         $[C_{ed}^{V,LR}]_{13jj}$ &     -1.13 &      -0.46 &        12.11 \\
         $[C_{eu}^{V,LL}]_{1311}$ &      2.28 &       0.92 &        49.59 \\
         $[C_{eu}^{V,LR}]_{1311}$ &      2.27 &       0.97 &       -23.60 \\
\hline

\multicolumn{1}{|c}{$[\mathcal{C}_{\ell q}^{(3)}]_{1333}(\Lambda) = 1 $} && & \\
         $[C_{ed}^{V,LL}]_{13jj}$ &      -1.14 &       -2.63 &        67.29 \\
         $[C_{ed}^{V,LR}]_{13jj}$ &      -1.14 &       -3.02 &        -9.50 \\
         $[C_{eu}^{V,LL}]_{1311}$ &       2.27 &        6.44 &       -60.48 \\
         $[C_{eu}^{V,LR}]_{1311}$ &       2.28 &        5.91 &        18.38 \\
\hline

\multicolumn{1}{|c}{$[\mathcal{C}_{\ell u}]_{1333}(\Lambda) = 1 $} &&&\\
         $[C_{ed}^{V,LL}]_{13jj}$ &  -  &       1.24 &        -58.33 \\
         $[C_{ed}^{V,LR}]_{13jj}$ &  -  &       1.27 &         12.77 \\
         $[C_{eu}^{V,LL}]_{1311}$ &  -  &      -2.56 &         47.26 \\
         $[C_{eu}^{V,LR}]_{1311}$ &  -  &      -2.43 &        -24.76 \\
\hline

\multicolumn{1}{|c}{$[\mathcal{C}_{\ell d}]_{1333}(\Lambda) = 1 $} &&&\\
         $[C_{ed}^{V,LL}]_{13jj}$ &      -1.14     &       -1.77  &        1.75 \\
         $[C_{ed}^{V,LR}]_{13jj}$ &      -1.14     &       -1.78  &        1.78 \\
         $[C_{eu}^{V,LL}]_{1311}$ &       2.28     &        3.57  &       -3.56 \\
         $[C_{eu}^{V,LR}]_{1311}$ &       2.28     &        3.51  &       -3.51 \\
\hline
                                      
\multicolumn{1}{|c}{$[\mathcal{C}_{e u}]_{1333}(\Lambda) = 1 $} &&&\\ 
         $[C_{de}^{V,LR}]_{jj13}$ &  -  &       1.28  &        -60.47 \\
         $[C_{ed}^{V,RR}]_{13jj}$ &  -  &       1.25  &         12.50 \\
         $[C_{ue}^{V,LR}]_{1113}$ &  -  &      -2.48  &         46.25 \\
         $[C_{eu}^{V,RR}]_{1311}$ &  -  &      -2.62  &        -25.80 \\

\hline

\multicolumn{1}{|c}{$[\mathcal{C}_{e d}]_{1333}(\Lambda) = 1 $} &&&\\
         $[C_{de}^{V,LR}]_{jj13}$ &  -1.14 &   -1.76 &   -1.75 \\
         $[C_{ed}^{V,RR}]_{13jj}$ &  -1.14 &   -1.75 &   -1.75 \\
         $[C_{ue}^{V,LR}]_{1113}$ &   2.28 &    3.48 &    3.47 \\
         $[C_{eu}^{V,RR}]_{1311}$ &   2.28 &    3.55 &    3.56 \\

\hline	 

\multicolumn{1}{|c}{$[\mathcal{C}_{q e}]_{1333}(\Lambda) = 1 $} &&&\\ 
         $[C_{de}^{V,LR}]_{jj13}$ &   -1.13 &   -0.49 &   -60.48 \\
         $[C_{ed}^{V,RR}]_{13jj}$ &   -1.13 &   -0.51 &    11.62 \\
         $[C_{ue}^{V,LR}]_{1113}$ &    2.28 &    1.04 &    47.08 \\
         $[C_{eu}^{V,RR}]_{1311}$ &    2.27 &    0.96 &   -23.86 \\

\hline
\end{tabular}
\end{adjustbox}

\caption{\small The impact of semileptonic operators on the low energy semileptonic $\Delta F=(1,0)$ Wilson coefficients 
(in $10^{-9}~ {\rm TeV^{-2}}$ units) at $\mu_{\rm low}=2$GeV is shown. The second column refers
to the values with only WET (QED+QCD) running, the third column refers to WET + SMEFT running
due to gauge interactions, and the fourth column refers to the full WET+ SMEFT
gauge + Yukawa running. Here the indices $jj=11$ or $22$ and $\Lambda=1$TeV. The SMEFT Wilson coefficients 
are set to 1 (in $\rm TeV^{-2}$ units) at the high scale.   }
\label{tab:wethadronic}
\end{center}
\end{table}

\section{Bounds on the Wilson coefficients}
\label{app:boundswcs}

In this section, we report the allowed ranges for the semileptonic Wilson coefficients $[c_X]_{ijkl}$ which 
are defined in \eqref{eq:smallc}.
Considering various 1-loop induced constraints, as discussed in Sec.~\ref{sec:observables}, ~ 
 the allowed values of the dimensionless parameters $[c_X]_{ijkl}$ are obtained by performing the fits. 
The results are presented in Tables~ \ref{tab:lq1}-\ref{tab:qe}. 
The uncertainties are also indicated at  $1\sigma$ level. In all cases, the cut-off scale $\Lambda$ is 
set to 3 TeV.

\begin{table}[H]
\renewcommand{\arraystretch}{1.0} \begin{table}[H] \centering \begin{tabular}{|lrrr|} \toprule \multicolumn{4}{|c|}{$\bar \ell_i \gamma^\mu \ell_j \bar q_k \gamma_\mu q_l$ }\\ \toprule  \multicolumn{2}{|c}{$\Delta F=(0,0)$} & \multicolumn{2}{c|}{$\Delta F=(1,0)$}  \\ \toprule 1111&$(-1.37 \pm 1.15).10^{0}$&1211&($1.64 \pm 0.62).10^{-2}$\\1122&$(6.94 \pm 1.68).10^{0}$&1222&($1.63 \pm 0.62).10^{-2}$\\1133&$(-7.42 \pm 3.71).10^{-1}$&1233&($2.42 \pm 0.91).10^{-3}$\\2211&$(8.77 \pm 6.07).10^{1}$&1311&$(3.99 \pm 1.03).10^{0}$\\2222&$(-6.32 \pm 1.46).10^{0}$&1322&$(7.48 \pm 2.74).10^{-2}$\\2233&($2.85 \pm 0.62).10^{0}$&1333&$(8.64 \pm 1.73).10^{-1}$\\3311&$(-2.68 \pm 1.82).10^{1}$&2311&($3.16 \pm 0.91).10^{0}$\\3322&$(-4.91 \pm 3.36).10^{1}$&2322&($1.19 \pm 0.46).10^{-1}$\\3333&$(-0.19 \pm 1.36).10^{0}$&2333&$(6.93 \pm 1.64).10^{-1}$\\\bottomrule \end{tabular} \end{table}
\caption{\small Allowed values of the Wilson coefficients $\wclow[(1)]{lq}{ijkl}$ based on the constraints 
discussed in the main text.}
\label{tab:lq1}
\end{table}

\begin{table}[H]
\renewcommand{\arraystretch}{1.0} \begin{table}[H] \centering \begin{tabular}{|lrrr|} \toprule \multicolumn{4}{|c|}{$\bar \ell_i \gamma^\mu \tau^a \ell_j \bar q_k \gamma_\mu \tau^a  q_l$ }\\ \toprule  \multicolumn{2}{|c}{$\Delta F=(0,0)$} & \multicolumn{2}{c|}{$\Delta F=(1,0)$}  \\ \toprule 1111&$(1.84 \pm 3.44).10^{-2}$&1211&$(5.75 \pm 2.15).10^{-3}$\\1122&$(2.08 \pm 1.85).10^{-1}$&1222&$(5.75 \pm 2.14).10^{-3}$\\1133&($-1.49 \pm 0.39).10^{0}$&1233&($2.07 \pm 0.77).10^{-3}$\\2211&$(7.41 \pm 5.86).10^{0}$&1311&$(3.99 \pm 1.31).10^{-2}$\\2222&$(-2.17 \pm 1.76).10^{-1}$&1322&$(7.56 \pm 2.74).10^{-2}$\\2233&$(4.01 \pm 2.53).10^{-1}$&1333&$(6.86 \pm 1.43).10^{-1}$\\3311&$(8.73 \pm 5.04).10^{-1}$&2311&$(3.13 \pm 1.09).10^{-2}$\\3322&$(1.71 \pm 1.01).10^{0}$&2322&($1.28 \pm 0.44).10^{-1}$\\3333&$(-0.10 \pm 1.07).10^{0}$&2333&$(5.40 \pm 1.30).10^{-1}$\\\bottomrule \end{tabular} \end{table}
\caption{Same as in Tab.~\ref{tab:lq1}, except for Wilson coefficient $\wclow[(3)]{lq}{ijkl}$.}
\label{tab:lq3}
\end{table}

\begin{table}[H]
\renewcommand{\arraystretch}{1.0} \begin{table}[H] \centering \begin{tabular}{|lrrr|} \toprule \multicolumn{4}{|c|}{$\bar \ell_i \gamma^\mu \ell_j \bar d_k \gamma_\mu d_l$ }\\ \toprule  \multicolumn{2}{|c}{$\Delta F=(0,0)$} & \multicolumn{2}{c|}{$\Delta F=(1,0)$}  \\ \toprule 1111&$(-1.49 \pm 2.26).10^{1}$&1211&($1.76 \pm 0.67).10^{-2}$\\1122&$(2.47 \pm 1.59).10^{1}$&1222&($1.76 \pm 0.67).10^{-2}$\\1133&$(-4.09 \pm 1.63).10^{1}$&1233&($2.24 \pm 0.84).10^{-2}$\\2211&$(-8.99 \pm 5.58).10^{1}$&1311&$(7.80 \pm 2.73).10^{-2}$\\2222&$(-7.04 \pm 2.16).10^{1}$&1322&$(6.94 \pm 2.59).10^{-2}$\\2233&$(4.34 \pm 2.12).10^{1}$&1333&$(7.77 \pm 1.94).10^{0}$\\3311&$(-0.22 \pm 6.83).10^{1}$&2311&$(6.36 \pm 2.25).10^{-2}$\\3322&$(-0.24 \pm 6.83).10^{1}$&2322&($1.18 \pm 0.43).10^{-1}$\\3333&$(-1.54 \pm 6.89).10^{1}$&2333&$(6.29 \pm 1.65).10^{0}$\\\bottomrule \end{tabular} \end{table}
\caption{Same as in Tab.~\ref{tab:lq1}, except for Wilson coefficient $\wclow[]{ld}{ijkl}$.}
\label{tab:ld}
\end{table}

\begin{table}[H]
\renewcommand{\arraystretch}{1.0} \begin{table}[H] \centering \begin{tabular}{|lrrr|} \toprule \multicolumn{4}{|c|}{$\bar \ell_i \gamma^\mu \ell_j \bar u_k \gamma_\mu u_l$ }\\ \toprule  \multicolumn{2}{|c}{$\Delta F=(0,0)$} & \multicolumn{2}{c|}{$\Delta F=(1,0)$}  \\ \toprule 1111&$(0.60 \pm 1.15).10^{1}$&1211&$(9.01 \pm 3.38).10^{-3}$\\1122&$(0.60 \pm 1.15).10^{1}$&1222&$(9.01 \pm 3.38).10^{-3}$\\1133&$(4.23 \pm 4.46).10^{-1}$&1233&($2.53 \pm 0.95).10^{-3}$\\2211&$(4.02 \pm 2.80).10^{1}$&1311&$(8.28 \pm 2.89).10^{-2}$\\2222&$(4.05 \pm 2.80).10^{1}$&1322&($3.42 \pm 0.84).10^{0}$\\2233&($-3.70 \pm 0.93).10^{0}$&1333&$(9.54 \pm 1.88).10^{-1}$\\3311&$(-0.52 \pm 3.44).10^{1}$&2311&$(6.45 \pm 2.50).10^{-2}$\\3322&$(-0.48 \pm 3.44).10^{1}$&2322&($2.72 \pm 0.74).10^{0}$\\3333&$(0.39 \pm 1.44).10^{0}$&2333&$(7.69 \pm 1.79).10^{-1}$\\\bottomrule \end{tabular} \end{table}
\caption{Same as in Tab.~\ref{tab:lq1}, except for Wilson coefficient $\wclow[]{lu}{ijkl}$.}
\label{tab:lu}
\end{table}

\begin{table}[H]
\renewcommand{\arraystretch}{1.0} \begin{table}[H] \centering \begin{tabular}{|lrrr|} \toprule \multicolumn{4}{|c|}{$\bar e_i \gamma^\mu e_j \bar d_k \gamma_\mu d_l$ }\\ \toprule  \multicolumn{2}{|c}{$\Delta F=(0,0)$} & \multicolumn{2}{c|}{$\Delta F=(1,0)$}  \\ \toprule 1111&$(1.75 \pm 2.83).10^{1}$&1211&($1.80 \pm 0.68).10^{-2}$\\1122&$(1.55 \pm 3.00).10^{1}$&1222&($1.80 \pm 0.68).10^{-2}$\\1133&$(2.15 \pm 3.03).10^{1}$&1233&($2.27 \pm 0.86).10^{-2}$\\2211&$(9.50 \pm 7.27).10^{1}$&1311&$(7.89 \pm 2.90).10^{-2}$\\2222&$(6.65 \pm 5.43).10^{1}$&1322&$(7.18 \pm 2.77).10^{-2}$\\2233&$(3.04 \pm 5.43).10^{1}$&1333&$(7.69 \pm 2.05).10^{0}$\\3311&$(-7.88 \pm 8.08).10^{1}$&2311&$(6.57 \pm 2.35).10^{-2}$\\3322&$(-7.92 \pm 8.08).10^{1}$&2322&($1.19 \pm 0.46).10^{-1}$\\3333&$(-9.74 \pm 8.16).10^{1}$&2333&$(6.15 \pm 1.56).10^{0}$\\\bottomrule \end{tabular} \end{table}
\caption{Same as in Tab.~\ref{tab:lq1}, except for Wilson coefficient $\wclow[]{ed}{ijkl}$.}
\label{tab:ed}
\end{table}

\begin{table}[H]
\renewcommand{\arraystretch}{1.0} \begin{table}[H] \centering \begin{tabular}{|lrrr|} \toprule \multicolumn{4}{|c|}{$\bar e_i \gamma^\mu e_j \bar u_k \gamma_\mu u_l$}\\ \toprule  \multicolumn{2}{|c}{$\Delta F=(0,0)$} & \multicolumn{2}{c|}{$\Delta F=(1,0)$}  \\ \toprule 1111&$(-0.78 \pm 1.35).10^{1}$&1211&$(8.71 \pm 3.31).10^{-3}$\\1122&$(-0.78 \pm 1.35).10^{1}$&1222&$(8.71 \pm 3.31).10^{-3}$\\1133&$(2.63 \pm 5.32).10^{-1}$&1233&($2.42 \pm 0.91).10^{-3}$\\2211&$(-4.43 \pm 3.51).10^{1}$&1311&$(7.73 \pm 2.61).10^{-2}$\\2222&$(-4.39 \pm 3.50).10^{1}$&1322&($3.30 \pm 0.84).10^{0}$\\2233&$(2.06 \pm 1.30).10^{0}$&1333&$(9.41 \pm 1.84).10^{-1}$\\3311&$(4.26 \pm 4.02).10^{1}$&2311&$(6.39 \pm 2.18).10^{-2}$\\3322&$(4.31 \pm 4.01).10^{1}$&2322&($2.65 \pm 0.66).10^{0}$\\3333&$(-1.44 \pm 1.57).10^{0}$&2333&$(7.76 \pm 1.67).10^{-1}$\\\bottomrule \end{tabular} \end{table}
\caption{Same as in Tab.~\ref{tab:lq1}, except for Wilson coefficient $\wclow[]{eu}{ijkl}$.}
\label{tab:eu}
\end{table}

\begin{table}[H]
\renewcommand{\arraystretch}{1.0} \begin{table}[H] \centering \begin{tabular}{|lrrr|} \toprule \multicolumn{4}{|c|}{$\bar q_i \gamma^\mu q_j \bar e_k \gamma_\mu e_l$ }\\ \toprule  \multicolumn{2}{|c}{$\Delta F=(0,0)$} & \multicolumn{2}{c|}{$\Delta F=(1,0)$}  \\ \toprule 1111&$(-1.98 \pm 2.76).10^{1}$&1112&($1.82 \pm 0.68).10^{-2}$\\2211&($-2.93 \pm 0.44).10^{1}$&2212&($1.83 \pm 0.68).10^{-2}$\\3311&$(-2.86 \pm 5.26).10^{-1}$&3312&($2.53 \pm 0.94).10^{-3}$\\1122&$(-8.69 \pm 7.11).10^{1}$&1113&$(8.26 \pm 2.96).10^{-1}$\\2222&$(5.22 \pm 4.87).10^{0}$&2213&$(6.90 \pm 2.68).10^{-2}$\\3322&$(-2.04 \pm 1.16).10^{0}$&3313&$(8.85 \pm 1.78).10^{-1}$\\1133&$(8.87 \pm 8.13).10^{1}$&1123&$(6.82 \pm 2.47).10^{-1}$\\2233&$(7.40 \pm 7.39).10^{1}$&2223&($1.16 \pm 0.44).10^{-1}$\\3333&$(1.32 \pm 1.52).10^{0}$&3323&$(7.30 \pm 1.67).10^{-1}$\\\bottomrule \end{tabular} \end{table}
\caption{Same as in Tab.~\ref{tab:lq1}, except for Wilson coefficient $\wclow[]{qe}{ijkl}$.}
\label{tab:qe}
\end{table}

\addcontentsline{toc}{section}{References}
\small
\bibliographystyle{JHEP}
\bibliography{Bookallrefs}
\end{document}